\documentclass[dissertation,letterpaper,12pt]{utthesis} 
\title{Building Containerized Environments for Reproducibility and Traceability of Scientific Workflows}	       	
\author{Paula Fernanda Olaya}                			
\copyrightYear{2020}            				
\graduationMonth{May}           				
\degree{Master of Science}	    			
\university{The University  of Tennessee, Knoxville}	
\usepackage{nomencl}                    
\usepackage{float}                      
\usepackage[numbers]{natbib}            
\usepackage{graphicx}					
\graphicspath{ {figures/}{figures/eps/}{figures/pdf/} }
\usepackage{fancyhdr}                   
\usepackage{url}                        
\usepackage[inactive]{srcltx}		 	
\usepackage{relsize}                    
\usepackage{booktabs}                   
\usepackage[config, labelfont={bf}]{caption,subfig} 
\usepackage{mathrsfs}                   
\usepackage[titletoc]{appendix}			
\usepackage{pdflscape}					
\usepackage{lmodern}  
\usepackage{amsmath}  
\usepackage{xcolor}   
\usepackage{listings}
\usepackage{comment}
\usepackage{multirow}
\usepackage{longtable}
\usepackage{tabularx} 
\usepackage{array}    
\usepackage{amsmath}

\fancypagestyle{mylandscape}{
	\fancyhf{} 
	\fancyfoot{
    \makebox[\textwidth][r]{
      \rlap{\hspace{.75cm}
        \smash{
          \raisebox{4.87in}{
            \rotatebox{90}{\thepage}}}}}}
}

\begin{document}
    \pagenumbering{alph} 
    \addToPDFBookmarks{0}{Front Matter}{rootNode} 
    \addToPDFBookmarks{1}{Title}{a} 
    \makeTitlePage 
    \pagenumbering{roman}
    \setcounter{page}{2}
    \makeCopyrightPage 
    %
    \addToPDFBookmarks{1}{Dedication}{b} 
    \chapter*{}
\begin{center}
{\centering \it To my family }
\end{center}  

    \addToPDFBookmarks{1}{Acknowledgments}{c} 
    \chapter*{Acknowledgments}
I wish to express my deepest gratitude to my advisor Dr. Michela Taufer, because as Malcolm Gladwell says in Outliers, \textit{"Success is not a random act. It arises out of a predictable and powerful set of circumstances and opportunities."}, and she personified those circumstances and opportunities at such a level that her guidance and encouragement are allowing me to achieve my dreams. 

I would like to recognize the invaluable and patient guidance of my mentor, Dr. Jay Lofstead. The support, advice and the brilliant ideas he provided throughout the past year allowed me to meet the goals of this work.

I would like to pay my special regards to the Singularity team, specially to Cedric (@cclerget) and Ian Kaneshiro, who responded to my questions and queries so promptly and instruct me into all the Singularity technicalities. Also special thanks to Andrew Younge for connecting me to the Singularity team. 

I wish to express my deepest gratitude to each member of GCLab, for their help, for listening to my presentation multiple times and giving me valuable feedback. Specially, to Silvina Caino-Lores for reading and reviewing it a thousand times;  to Dylan Chapp and Michael Wyatt, for being my big brothers in the past two years. 

On a personal level, I wish to acknowledge the support and great love of my family, my mother, Nancy; and my father, Armando. They kept me going on and this work would not have been possible without their and my friends emotional input and their cheers of encouragement.

    
    \addToPDFBookmarks{1}{Abstract}{e} 
    \chapter*{Abstract}\label{ch:abstract}
Scientists use simulations to study natural phenomena, and trusting the simulation results is vital to the integrity of scientific discovery. To trust results, we must ensure the simulations’ reproducibility, replicability, and traceability through the annotation of simulation's executions. The annotation allows us to build a record trail of data moving within a given simulation workflow. Past efforts advocated for the need to build record trails at the system level but a key hindrance of these approaches was the limits of the system technology. The evolution from virtual machines to containers has opened new opportunities for system-level solutions.
In this work, we propose an operative system-level solution that leverages the intrinsic characteristics of containers (i.e., portability, isolation, encapsulation, and unique identifiers) to annotate workflows and capture their metadata. Our solution enables transparent and automatic metadata collection and access, easy-to-read record trail, and tight connections between data and metadata. We build a prototype of a containerized environment which encapsulates each component of a scientific workflow (i.e., data and applications) in individual containers. Our prototype implementation features zero-copy data transfer between containers, requires no modification of the underlying applications, and automatically links the metadata to the workflow. We assess the effectiveness of our prototype for four increasingly complex workflows, ranging from simple visualization applications such as, gnuplot to machine learning applications such as KKNN and random forest; and show that we are able to build workflow record trails at the OS-level for all four scenarios in an automatic, easy-to-read, and with a tight connection between data and metadata. We measure the costs of our containerized environment in terms of time and space. We observe that time overhead associated with the containerization becomes tolerable when the workflows have larger size and longer run time applications. We also observe that the space overhead is driven by the OS, software stack, and file system. Our containerized environment addresses metadata from OS system-level by leveraging cutting edge container technology to provide a complete, transparent and automatic collection and management of workflow metadata.  

    \addToPDFBookmarks{0}{Table of Contents}{f}
    \tableofcontents 
    \listoftables 
    \listoffigures 
   
    \newpage
    \pagenumbering{arabic}
    \setcounter{page}{1}
    \include{front-matter}
    \chapter{Thesis Overview}


Scientists execute multiple simulations to study natural phenomena. Trusting the simulation results is vital to develop sciences in any field. To trust the results, we must ensure the simulations’ replicability, reproducibility, and traceability through the annotation of simulation's executions. We consider the scope of the words 'replicability' and 'reproducibility' as defined in \cite{reproducibility}, and 'traceability' as defined in \cite{Digi_Trace}.

The annotation allows us to build a record trail of data moving across a simulation workflow. This record trail seeks to answer simple questions like: What is the origin of the data? Has it been modified during the simulation? How? What ontology or rules were used to generate some portions of the data? The answers to these questions are included in the metadata file which should provide enough and understandable information so in case it is shared among collaborators, the simulation can be executed and similar results are obtained. 

Besides, building the record trail of the workflows and add it to the metadata, we adopted the idea from Muniswamy-Reddy et al~\cite{pass}, who already in 2005 advocated for the need to address metadata from within the system-level. The system-level approach offers full visibility into system-wide behavior, and as workflows are executed and data flows though the operative system and the storage system, it allows one to have transparency and completeness to build, collect and access the metadata. In 2005, the system technology included virtual machines which are heavy and sometimes slow, however, in 2013 container technology were introduced. Containers are lighter and faster software systems. This evolution of system-technology has opened new opportunities for system-level solutions. 

This work is a collaboration with Sandia National Lab where we have been building upon this vision that
seeks to leverage cutting edge container technologies to address metadata from the OS level and overcome the flaws from the current systems. 
Containers offer multiple functionalities, such as, encapsulation, isolation, portability,  and unique identifiers. 
Encapsulation allows the addition of multiple data and applications in a single object (a container).
Isolation limits what the containers can see, providing their own isolated instance. Portability ensures that applications run on the containers can be reproduced on any platform
independent libraries or packages on it.  Lastly, each container has assigned a universally unique identifier (UUID).  

Acknowledging the functionalities that container technology offer, we build a prototype of an application-agnostic containerized environment. Every workflow component in our environment is encapsulated into their own system generated containers. Our environment
builds the record trail from the workflow and collects this information in the metadata. 

Our prototype implementation features zero-copy data transfer between containers, requires no modification of the underlying applications, and automatically links the metadata to the workflow.  

We assess the effectiveness for four representative workflows comprising two applications: a visualization applications such as, gnuplot, and two machine learning applications such as Weighted k-Nearest Neighbors (KKNN) and random forest. 

We show that we are able to build workflow record trails at the OS-level for all four scenarios automatically. Our prototype enables transparent and automatic metadata collection and access, easy-to-read record trail, and tight connections between data and metadata.  

We measure the costs of using our containerized environment in terms of time and space. We observe that time overhead associated with the containerization becomes tolerable when the workflows have larger size and longer run time applications. We also observe that the space overhead is driven by the OS, software stack, and file system. 

Our containerized environment addresses metadata from OS system-level by leveraging cutting edge container technology to provide a complete, transparent and automatic collection and management of workflow metadata. 

\section{Thesis statement}
In this thesis, we claim that there is a need for tools that annotate the workflow execution to ensure reproducibility, replicability, transparency and traceability of simulations. 

To validate the thesis statement, we:
\begin{itemize}
\item Analyse relevant works in the scientific literature related to the replicability, reproducibility transparency and traceability of scientific workflow, in order to find the limitations in state-of-the-art solutions. 
\item Design and implement a environment that annotates workflow executions, assembling a record trail with information of the workflow components that produced the new data. 
\item Enhance our environment to address annotations at OS-level by leveraging container technology.   
\end{itemize}

\section{Contributions}
The contributions of the thesis are as follows:

\begin{itemize}
\item  A novel solution to address metadata at OS level by leveraging container technology.  
\item A prototype of an application-agnostic containerized environment that enables transparent and automatic metadata collection and access, easy-to-read record trail, and tight connections between data and metadata.
\item An augmentation of containers features, to transfer data between containers and to support workflow record trail assembly. 
\end{itemize}

\section{Organization}

The remainder of this thesis is structured as follows. 
Chapter~\ref{ch:relatedwork} studies five key tools, characterized by their ability to collect metadata for executions in terms of software system, workflow, and application metadata. We critically assess if each tool fulfills the requirements to be a complete system-level solution implemented to collect and manage workflow metadata. 
Chapter~\ref{ch:methodology} explains the considerations made on each step of the development of our agnostic-application containerized environment.
Chapter~\ref{chap:implementation} presents the design and implementation with Singularity container technology of a diverse set of scenarios on which we assess the capability of our environment to successfully produce a result.
Chapter~\ref{ch:costs} provides the cost in terms of time and space to deploy our containerized environment.
Chapter~\ref{ch:conclusion} summarizes our work and future committed research.

    \chapter{Related Work} \label{ch:relatedwork}

\begin{figure}[!ht]
\centering
 \includegraphics[width=1.0\textwidth]{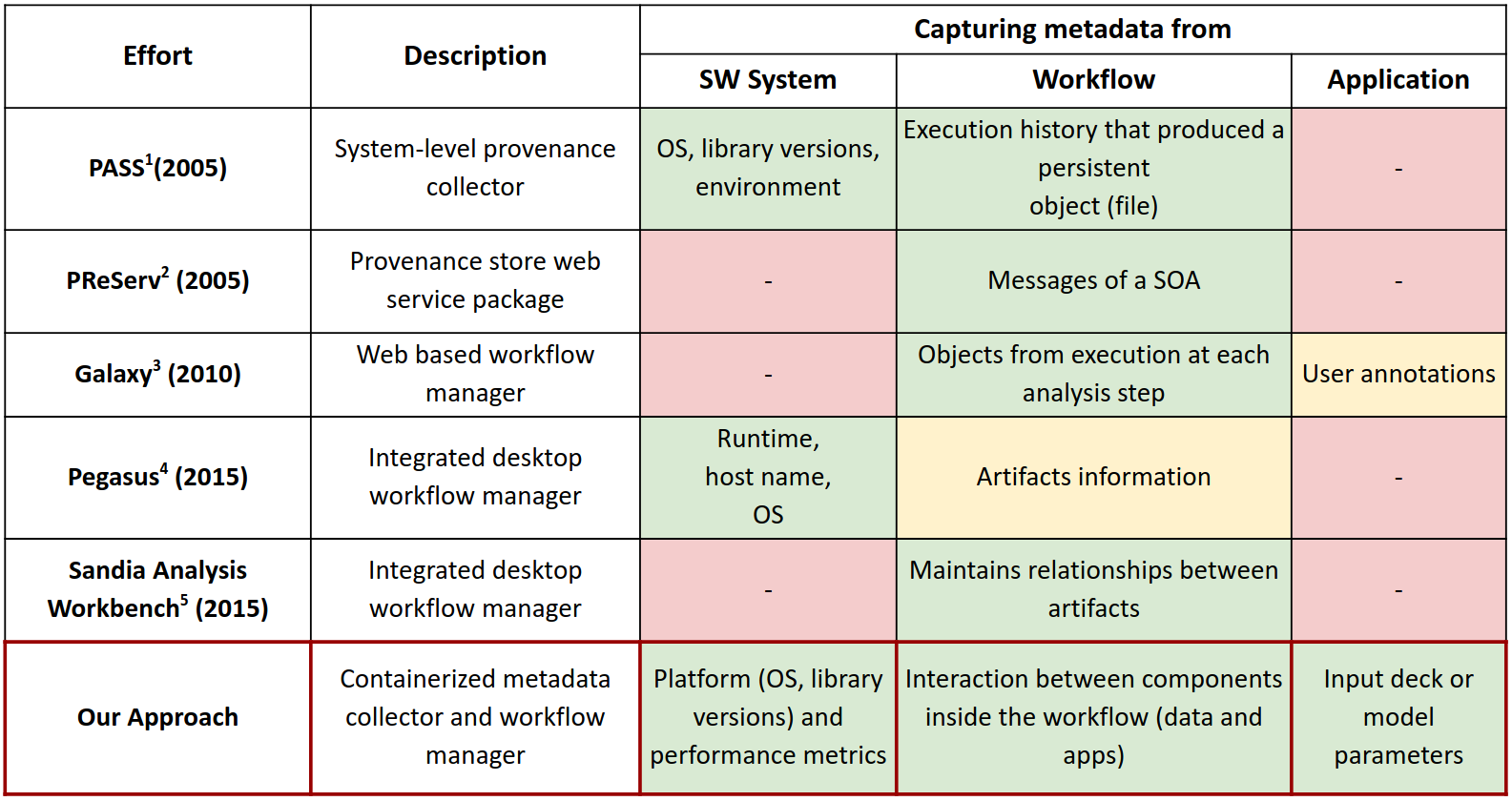}
 \caption{Overview of five milestone tools and our solution based on the information included in the metadata.}
 \label{fig:metadatacontent}
\end{figure}

Scientists need tools that allow them to build record trails of executions in a general manner, independently from the type of software environment, workflow configuration, and application. One way to do so is to leverage system functionality (e.g., at the OS level, or at storage system level). Efforts to collect and manage record trails through system functionality have a key precursor in PASS (or Provenance-Aware Storage Systems)~\cite{pass}. Other tools have been built with some aspects advocated by PASS. Figure~\ref{fig:metadatacontent} shows five milestone tools, including PASS, characterized by their ability to collect record trails for executions in terms of software system, workflow, and application metadata. 

Starting from PASS, we discuss how these tools have evolved to included metadata extraction and what components support what features of the metadata extractions.

\section{Provenance-aware storage systems}

\textbf{PASS} advocated for the need to address metadata from within the system level already in 2005. In~\cite{pass}, the authors claimed \textit{“We believe that as all information flows through the operating system, the operating system in cooperation with the storage system should be responsible for the collection and management of provenance.”} 

PASS captures the provenance or record trail of workflow data in a DAG. PASS also includes a complete description of the hardware platform on which outputs are generated with a description of the operative system and system libraries, as well as the process environment. However, PASS fails to capture full information for the application execution, including limited metadata such as random number generator seed to make pseudo-random computation repeatable and some user-based annotations of the files.

\section{Web service packages}

Provenance store web services that build on PASS principles have among the first precursors \textbf{PReServ}~\cite{preserve}. PReServ is a software package that contains a set of interfaces for recording and querying provenance. It also has a set of Java libraries for easily accessing the interfaces, and a handler for recording message exchange for web services. PReServ works across a spectrum of applications and cross-platform: it was tested on Mac OS X, Windows, and Linux). It also has a flexible architecture allowing users to add or modified functionalities. 

PReServ records process documentation in a generic manner for Service Oriented Architecture (SOA). Therefore, it is able to capture workflow information about the content of messages along with the relationships between services and the trace of how a process led to a results, allowing users to distinguish specific steps of a process from the whole process. On the other hand, PReServ does not capture software system information (e.g., platform where they run or the software system), neither scientific application information (e.g., information related to the scientific model). 

\section{Workflow managers}

There is rich and diverse spectrum of workflow managers that integrate some of the features and principles promoted by PASS. 

Among these solutions, \textbf{Galaxy} is an example of a comprehensive approach for supporting accessible, reproducible, and transparent computational research in the life sciences~\cite{galaxy}. Galaxy is an open web-based genomic workbench for computational analysis. Specifically, it is a collaborative environment for performing complex analyses in genomics, supporting automatic and unobtrusive provenance tracking of precise computational details, intent, context, and narrative.

Galaxy captures automatic workflow metadata (i.e., provenance, inputs, parameters, and output) for each tool used at each analysis step. It also offers to group different steps; the user has the flexibility to associate the steps by preference. 
Galaxy captures parameters that affect an application simulation sufficient to repeat the analysis including user annotation (i.e., descriptions or notes from critical information that is not automatically captured). Galaxy does not include any software system  information.

\textbf{Pegasus} is among general purpose workflow manager for science automation~\cite{pegasus}; it executes, maintains and administrates a broad set of scientific workflows, including {\bf add examples and citations}. Pegasus targets large-scale workflows running in distributed environments. Pegasus separation of workflow description from execution environment description, makes this workflow manager portable across executions environments; optimizations can be done at compile, or run time, or both. 

Metadata about the status of the software system while executing an application is obtained through job logs and includes aspects such as job arguments, start and duration times, jobs' stdout and stderr, job environment and machine information (e.g., architecture, operating system, number of cores on the node, and available memory). Pegasus also captures information about the workflow in a DAG format including the workflow jobs and their input tasks, record trails of process, local workflow execution engine logs, and pre-execute and post-execute scripts associated with a workflow.
However, Pegasus does not currently provide in information about applications' input deck.

Last, the \textbf{Sandia Analysis Workbench or SAW}~\cite{saw} is a family of desktop applications developed to provide an integrated interface that enables scientists to access, monitor, and manage their scientific simulations. 
SAW does not include information about the status of the software system while executing a workflow. 
The data management component of SAW only maintains relationships between artifacts but it does not store the complete record trail, missing to record the application and its parameters. On the positive side, SAW offers scientists with a powerful interface for creating a Plugin which can include user annotations.

\section{Overview of the tools features}
\begin{figure}[!ht]
\centering
 \includegraphics[width=1.0\textwidth]{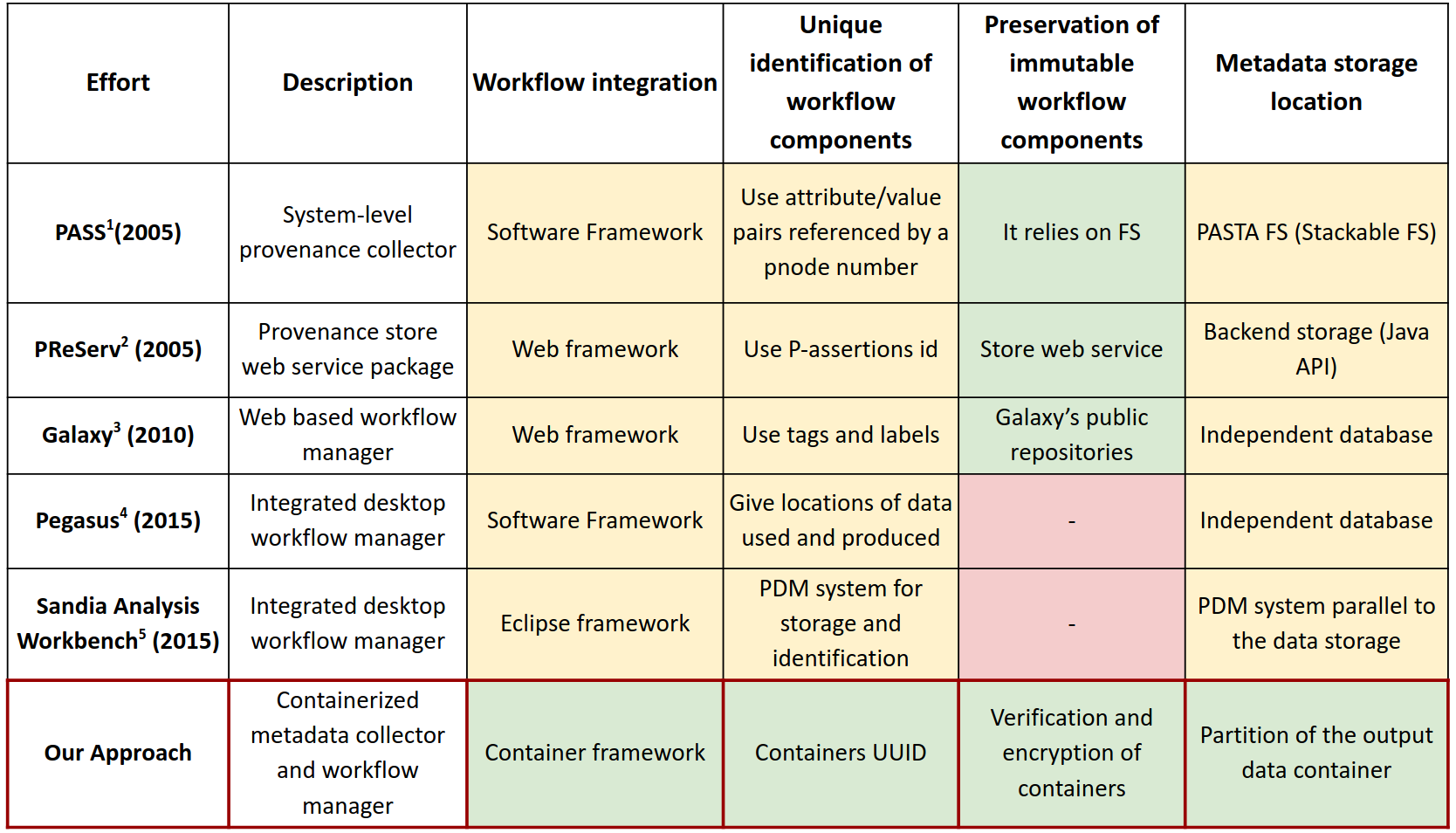}
 \caption{Overview of five milestone tools and our solution based on the framework plugin, the identification and preservation of workflow components and the metadata storage. }
 \label{fig:othercolumns}
\end{figure}

In Figure~\ref{fig:othercolumns} we present an overview, of each of the tools, regarding the implementation of the listed features: integration of the workflow to their frameworks, identification and preservation of the workflow components, and the metadata storage.

\subsection{Workflow integration}
\textbf{PASS } approach was implemented in Linux 2.4.29, and since it is a system-level solution it did not require modifications to the application's format, the user only needed to run the command line in the terminal and given the file system modification made by PASS, the metadata is collected.

\textbf{PReServ} is implemented by using Java and a set of Plug-Ins which add a specific functionality to the software. Even when it is intended for SOA application, command line programs can be wrapped by a script that was submitted p-assertions to the Provenance Store. 

\textbf{Galaxy} is intended to be used even for users that do not have a program. It has a repository of applications and data where the user can select and play around to get their results. Furthermore, Galaxy also allows integration of new programs (any piece of software in any language) and new data. To add the new program to their framework, it is required by the developer to write a configuration file that describes how to run the simulation, including detailed specification of input and output
parameters. However, when adding a new application to their framework, it offers less flexibility and some of the feature cannot be used. Another option is to use Galaxy’s interactive, graphical workflow editor, which supports the creation from scratch of workflows. 

In \textbf{Pegasus} users can interact with their framework  via the command line and API interfaces through portals or trough application-specific composition tools. Pegasus workflows are represented as Directed Acyclic Graphs (DAGs), where the nodes represent individual computational tasks and the edges represent data and control dependencies between the tasks. Therefore, in order to have a compatible workflow to be executed, the user must instrument their code with the specific functions  to generate the workflow in the right format. Pegasus offers easy API's in Python, Java and Perl that convert the application into the correspondent format.

\textbf{SAW} is based on the Eclipse framework, where they use a data-driven or declarative approach in which the syntax of a code is described in a data file, and graphical interfaces are generated at run time from that description. SAW includes several code generators that help the user to modify from their own code to the needed format. 

\subsection{Identification and preservation of workflow components}
\textbf{PASS} reference each object from the workflow that was used to generated the new data by a unique identifier called a \textit{pnode number}. Even when it claims it to be unique, this number can be reused in between executions of different workflows. To preserve immutable workflow components, PASS relies on the memory system, which implies that the elements can be easily replace and erased. Also their collector, has a method for duplicates elimination, so once you re-execute your workflow, the previous values form the components are not taken in account anymore.

\textbf{PReServ }follows the P-assertion Recording Protocol (PReP) where there is an \texttt{identifier} type which contains a set of ids that link a service assertion to a p-structure. The p-structure is represented as as XML Schema. There is no preservation of the workflow components.

\textbf{Galaxy} supports tagging, adding words or labels to identify and describe an item. This is useful for categorizing and searching for scientific applications, specifically web applications. These tags and labels are not unique across workflow and executions. Regarding the preservation of components, Galaxy allows sharing the models where every workflow component and its metadata can be published to Galaxy's public repositories. Besides, the shared or published items can be imported and reused.

\textbf{Pegasus} identifies data components inside the workflow by the location. We could catalogue that Pegasus lacks a depth method to identify and preserve workflow components in order to be replicated, reproduced and traceable.

\textbf{SAW} defines the Workbench Server to store data in a commercial product data management (PDM) system. This PDM provides versioned storage and it identifies the components based on naming or labeling the files generated. They explain that every model is stored in a \textit{project} which has associated a team that can access the files.

\subsection{Metadata storage location}
\textbf{PASS} has a storage layer composed of a stackable file system, called PASTA to store its metadata. PASS claims the need to easily access metadata, that is why a database library is used to store and index the metadata. The database (Berkeley DB) provides fast, indexed lookup and storage for key/values pairs. Even when the data is not in the same file system as its metadata, they link and the user is able to map to the specific \textit{pnode number} of the file.

Backend storage is where p-assertions are finally stored in the Provenance Store. \textbf{PReServ} has three backend implementations: 1) the file system backend which stores p-assertions in a hierarchy of directories directly on the file system. This backend is particularly useful in debugging as it allows the contents of the Provenance Store to be inspected by the user, 2) the in-memory backend stores system uses hashtables and vectors to model the p-structure, and 3) the database backend  that is critical because a large amount of p-assertions can be stored using it.

\textbf{Pegasus} has a structure called The Mapper which is in charge of assembling the metadata, also logs this information to a local file in the workflow directory. With the detailed logs, Pegasus can generate the provenance of the results, referring to the original workflow. Then, the database stores both performance and the file created by The Mapper. It also sends notifications back to the user notifying events such as failures, success, and completion of tasks, jobs, and workflows.

\textbf{Galaxy} and \textbf{SAW} use an independent database for storing the metadata.

\addtocontents{toc}{\protect\setcounter{tocdepth}{0}}
\section{Summary}
In this Chapter we were able to observe that there is not complete system-solution developed to collect and manage workflows metadata that includes the record trail. None of the studied tools have a complete metadata content in terms of the software environment, the workflow execution, and the scientific model. Besides, we studied some features, concluding that all the tools require code or format modifications to integrate the workflow to their framework; there is a lack of unique identification and preservation of workflow components; and most of the tool, used independent data bases to store their metadata. 
In the last row from Figure~\ref{fig:metadatacontent} and Figure~\ref{fig:othercolumns} we present how our overarching vision seeks to satisfy all the requirements in order to be a complete OS-level solution. In the next Chapter~\ref{ch:methodology} we explain step by step the process and considerations to build this tool.
\addtocontents{toc}{\protect\setcounter{tocdepth}{2}}
    \chapter{Building Containerized Environments} \label{ch:methodology}

We describe our process for building general, application-agnostic containerized environments by encapsulating each component of a workflow (i.e., input and output data, applications) into individual containers and capturing the workflow record trail (i.e., metadata) at run time. To this end, we need to meet two requirements: 
\begin{itemize}
    \item Create an application-agnostic containerized environment
    \item Define the general metadata specifications
\end{itemize}

In this work we focus on the development of the first requirement and move the first steps towards the second requirement. 

\section{Stages for building containerized environments}
\label{application-agnostic}

Our solution to create a general application-agnostic containerized environment consists of eight stages. First, we leverage container technologies to host workflow executions. Second, we decouple a workflow into components (i.e., data and applications). Third, we encapsulate a single component in a single container (one to one). Fourth, we connect the containerized components to rebuild the original workflow. Fifth, we execute the containerized workflow. Sixth, we capture containers’ metadata (fine-grained information). Last, we build the record trail (workflow metadata) backwards from the fine-grain information and the communication patterns from that execution. We describe each step in detail below.

\section{Leveraging container technology}
\label{leveragecontainers}


Leveraging container technologies raises two key questions: (a) what are the benefits of container technologies and (b) which specific container technology better fits our needs in high performance computing. 

To answer the first question we consider the properties that container technology offers. The first property is \textbf{portability}, defined as the the ability to have immutable applications machine-agnostic. The second property is \textbf{isolation}, where through \texttt{namespaces} one can group processes and limit what they can see, providing their own isolated instance. The third property is the \textbf{encapsulation} of single objects, which is possible through the layered design of containers. The last property is the \textbf{unique identification} where for each container is guaranteed to have a universally unique identifier (UUID).

To answer the second question we notice the functionalities of Singularity\cite{singularity}. First, Singularity is a container technology that does not require administrative privileges. This property makes Singularity compatible with HPC architectures. Singularity  also provides users with mobile and reproducible software stack , and a strong security model. Other container technologies such as Docker~\cite{docker} do not support all this rich spectrum of properties.

\begin{figure}[!ht]
\centering
\includegraphics[width=7cm,height=2cm]{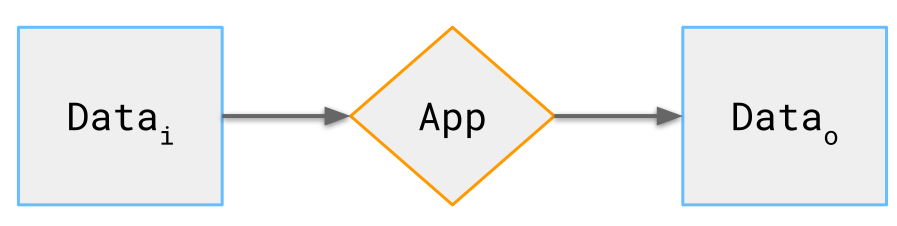}
\caption{Simplified workflow a single input file ($Data_i$), an application ($App$), and an output  file ($Data_o$).}
 \label{fig:simplified}
\end{figure}

Figure~\ref{fig:simplified} shows the three components of a simplified workflow that we will encapsulate in Singularly containers to build our containerized environment. This simple workflow includes a single input file ($Data_i$), an application ($App$, and an output  file ($Data_o$). 

\section{Decoupling components of the workflow}

The taxonomy of our simplified workflow consists of data (input and output) and applications. Data is the compilation of individual units of information. It can be represented in two ways: input data and output data. The \textbf{input data} serves as input to the program or application. While the \textbf{output data} is generated or modified by the program or application. With respect to the \textbf{application (App)}, it is a program that executes a transformation on the data. For our purpose the applications we target are non-distributed. As future work we intent to work with applications that require large resources requirement. 
 
\begin{figure}[!ht]
\centering
\includegraphics[width=7cm,height=1.5cm]{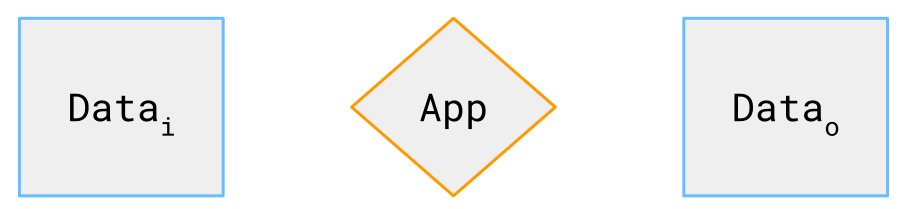}
\caption{Decoupling each of the components from our simplified workflow.}
\label{fig:decoupled} 
\end{figure}

In Figure~\ref{fig:decoupled} we show three components of our simplified workflow.

\section{Encapsulating components in containers}\label{sec:encapsulate}

\begin{figure}[!ht]
\centering
\includegraphics[width=8.7cm,height=3cm]{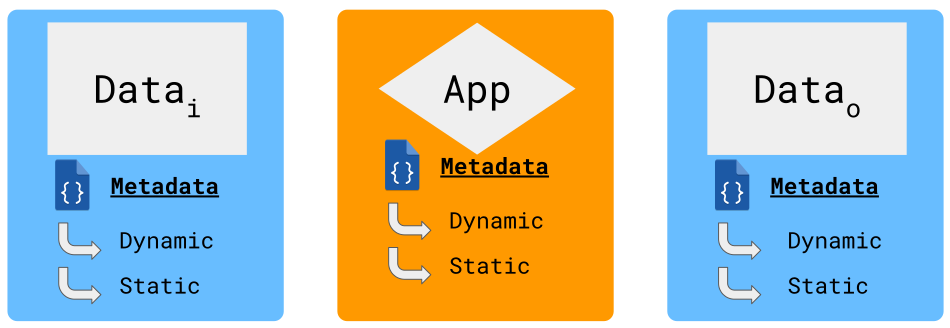}
\caption{Encapsulation process of each of the components from the simplified workflow. }
 \label{fig:containers}
\end{figure}

In this Section we encapsulate each of the decoupled workflow components. In essence, our environment will have two types of containerized workflow components: data container and application container. In Figure~\ref{fig:containers} we show the encapsulation of each of the component into individual containers. 

For each of the two components, we need to address different considerations, given that the components have to be correct wrapped in order to be correct connected for execution. 

\subsection{Data container}
To create this container we compress the data and added as a single and independent partition. As a first consideration to maintain the data, we need to find the right file system. It must be a read and write file system because the data will be used (read) and transferred (write) across the workflow.

\subsection{Application container}
Regarding the app container, we need to have all the software environment dependencies to be able to execute the application. The user must be able to define the operative system of preference, the packages and the libraries in the specific versions.

That is why we use the Singularity feature to  build it from a recipe. This recipe allows the user to define all the software-system specifications. In the end our application container has a system single and independent partition.

\section{Connecting containers}
In this Section we explore the current way of sharing data between containers in Singularity \textbf{(two-copy data transfer)}, and how together with the Singularity team we were able to expand its original feature and implement a direct transfer copy or \textbf{(zero-copy transfer)}. 

\subsubsection{Two-copy data transfer}
\begin{figure}[ht]
\centering
\includegraphics[width=8cm,height=5cm]{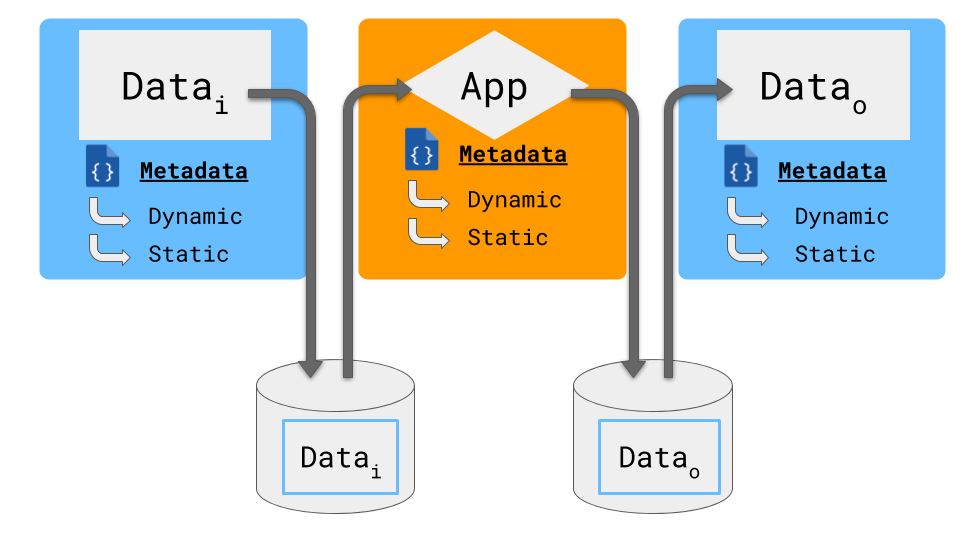}
\caption{Two-copy data transfer with a simplified workflow. }
 \label{fig:twocopy}
\end{figure}

Singularity lacks a method to map directories from multiple containers directly. However, it allows to map directories on your host system within your container to read and write data. Based on the workflow configuration shown in \ref{fig:twocopy} we have four connections: (1) to move the data from the input container to the host system, (2) from the host to the application container, (3) from the application container to the host, and (4) from the host to the output container.

Figure~\ref{fig:twocopy} presents the current method from Singularity to transfer data between containers.

\subsubsection{Zero-copy data transfer}
\begin{figure}[ht]
\centering
\includegraphics[width=8.5cm,height=3cm]{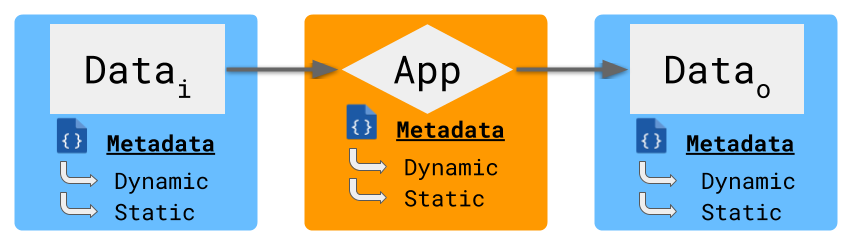}
\caption{Zero-copy data transfer with a simplified workflow. }
 \label{fig:zerocopy}
\end{figure}

Figure~\ref{fig:zerocopy} shows how the feature to transfer data between container is expanded avoiding to go through host to share data, we call it zero-copy data transfer.

We collaborated with the Singularity team to add the feature to map directories directly within containers. We can connect as many containers as needed in execution. We present the implementation of this in Subsection~\ref{ssec:implementation_zero-copy}

\section{Executing and capturing container's metadata}
\begin{figure}[ht]
\centering
\includegraphics[width=8.5cm,height=3cm]{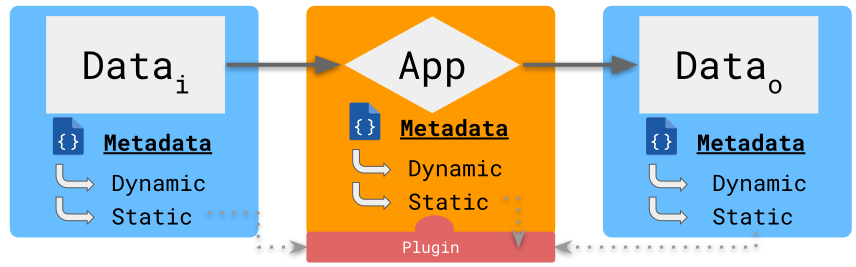}
\caption{Singularity plugin which interacts with all the components from the workflow and extracts their static metadata to build the record trail. }
 \label{fig:plugincapture}
\end{figure}

After the workflow is containerized and optimally connected, we execute it. 
As observe in Figure~\ref{fig:plugincapture}, we developed a Singularity plugin which is a package dynamically loaded that interacts with more complex subsystems of Singularity at run time. For our purpose, it interacts to all the containers used in execution to extract their static metadata.

This plugin creates metadata file. It includes the information captured from the containers, in particular their static metadata, such as, name, UUID, creation date, and modification date. 
 
With this information we are able to build the record trail (workflow metadata) backwards and determine the communication patterns from that execution.

\section{Locating the metadata}
Once the file with the record trail of data is created we need to attach it to the workflow. The Singularity plugin has the functionality of doing it. 
We explored two allocation methods: (1) metadata in isolation, and (2) metadata container-collocated. In the first method, the file is  added as a partition to an new container. The second method allocates the metadata as a new partition from the output data container. Because of the consistency and relation between the data and the metadata, our environment uses the second method. \\

\subsection{Metadata in isolation}
\begin{figure}[ht]
\centering
\includegraphics[width=8.5cm,height=4.7cm]{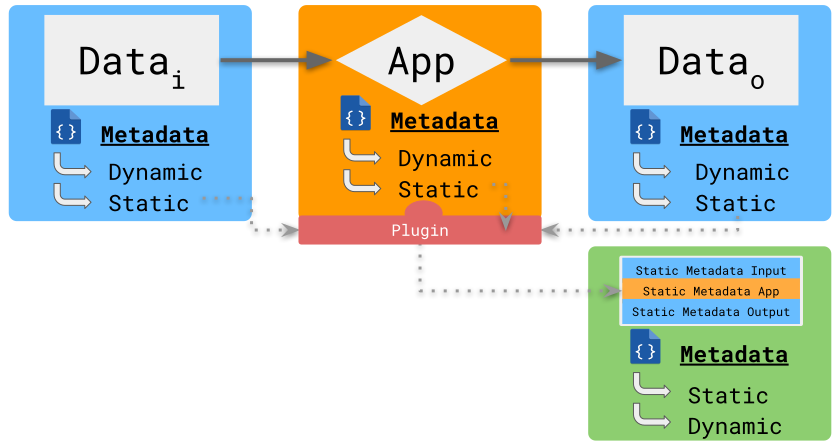}
\caption{Allocation of the metadata in an isolated container.}
 \label{fig:isolated}
\end{figure}

This method is shown in Figure~\ref{fig:isolated} where  the file is added as a partition to an new extra container. The file is compressed and added as a single and independent partition to the \textit{metadata container.}

\subsection{Metadata container-collocated}
\begin{figure}[!ht]
\centering
\includegraphics[width=8.5cm,height=3.2cm]{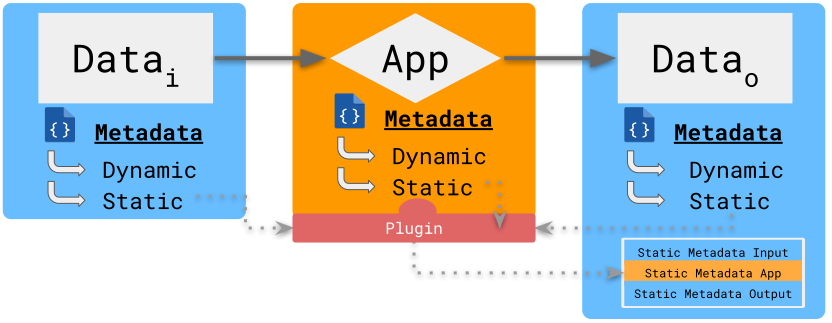}
\caption{Allocation of the metadata in the output data container as a JSON Generic partition. }
 \label{fig:collocated}
\end{figure}

Figure~\ref{fig:collocated} presents the second method where the data is allocated in the output data container. The Singularity plugin access to the output data container and creates a new partition. The metadata with the record trail is attached to that new partition. 


In this Chapter we explained the considerations made on each step of the development of our agnostic-application containerized environment. In the next Chapter~\ref{ch:implementation} we will present the details of implementing it with the Singularity container technology.

    \chapter{Implementing Containerized Environments}
\label{ch:implementation}

We present the design and implementation of a diverse set of workflow scenarios on which we assess the degree that our environment is successful in producing a result.

\section{Workflow scenarios}
\label{}

\begin{figure}[!ht]
\centering
    \includegraphics[width=8.3cm,height=5.5cm]{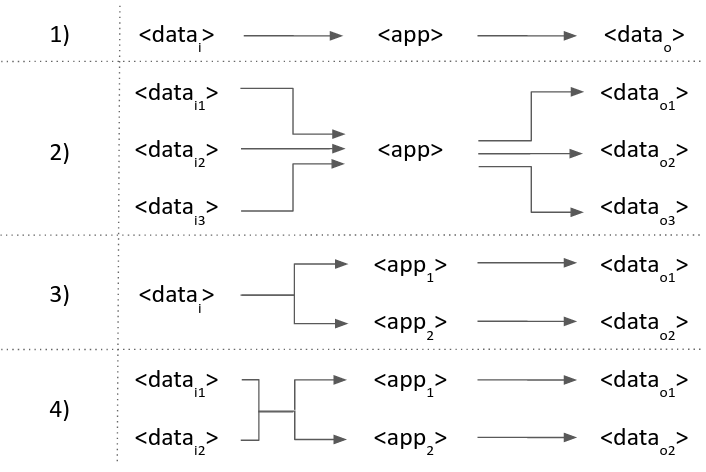}
    \caption{Configuration of assessed workflows.}
    \label{fig:generalworkflows}
\end{figure}

We consider four general workflows: (a) a simple application with single input and single output; (b) an application with three inputs and three outputs; (c) two application with a single input and two outputs; (c) two applications with two inputs and two outputs. 
Figure~\ref{fig:generalworkflows} presents the four general workflows.  

\begin{figure}[ht]
\centering
 \includegraphics[width=9.5cm,height=5.3cm]{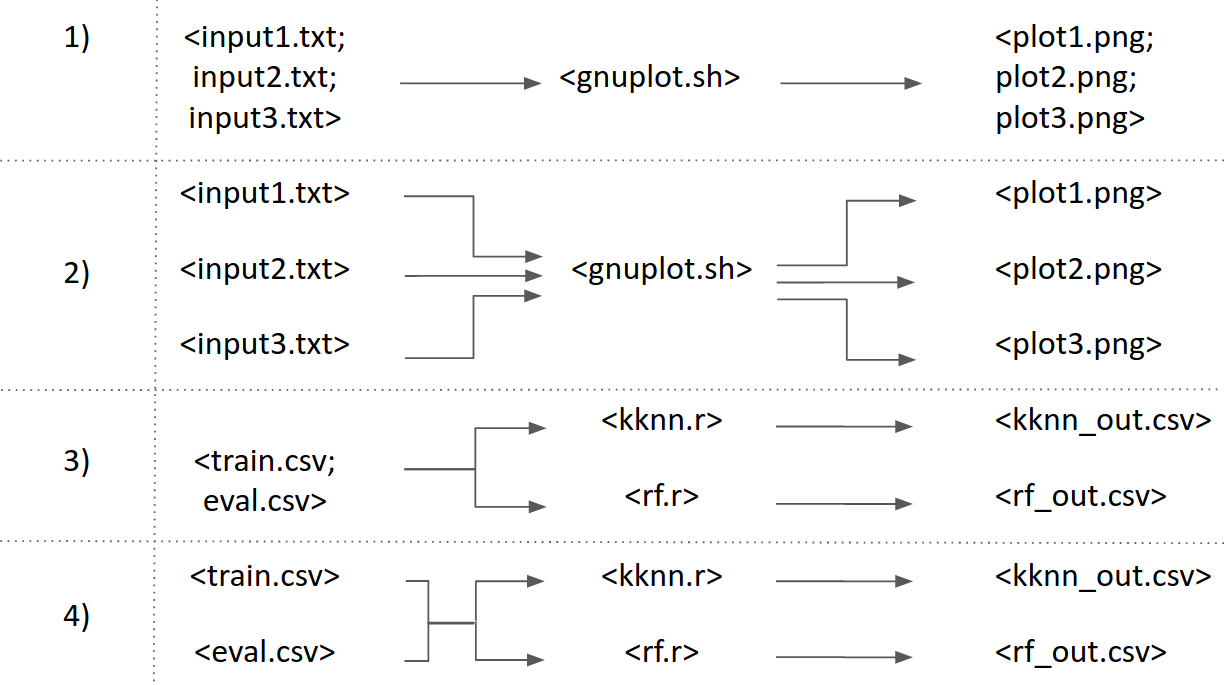}
 \caption{Specific scenarios with a gnuplot benchmark and two ML models.}
 \label{fig:selectedscenarios}
\end{figure}

For each of the four general workflows, we define a specific workflow scenario with real benchmarks and applications. Figure~\ref{fig:selectedscenarios} shows the specific workflows. The first two scenarios use a visualization application (i.e., gnuplot benchmark). The input files for the first and second workflow are text files. The application plots each of the input files. The third and fourth scenarios use two machine learning applications (i.e., Weighted k-Nearest Neighbor Classifier (KKNN) and random forest). Their input files are \texttt{CSV} files. The two applications generate predictions in a csv file as well. 

\section{Scenario 1: Gnuplot benchmark with single input}\label{s1}

This scenario consists of a single input, a single application executed, and a single generated output. 
The input is a directory with three text files that include the data points and specifications of a given plots.
The application plots in output from the input text files using \textit{gnuplot}. The output is a directory with the compilation of the generated plots. 

Here we are able to observe our environment's data versatility, where we can maintain and transform from \texttt{.txt} files to \texttt{.png} figures.

\subsection{Singularity implementation}
We present a detailed development and implementation of our environment for scenario 1. 
We follow the same process as in Chapter~\ref{ch:methodology}.
\subsubsection{Decoupling components of the workflow}
\begin{figure}[h]
\centering
\includegraphics[width=7cm,height=3.2cm]{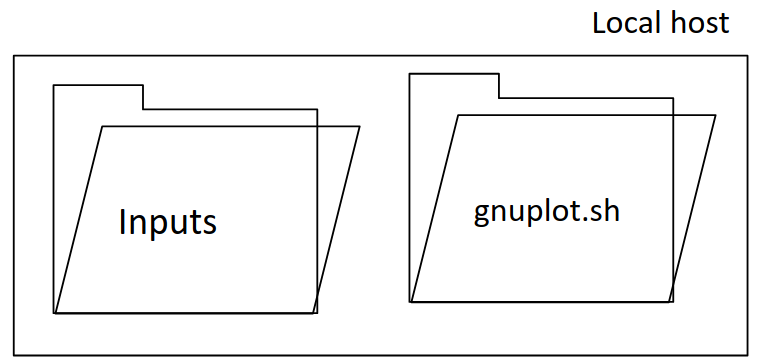}
\caption{Input data and executable from the original workflow. }
 \label{fig:s1_decoupled}
\end{figure}

In Figure~\ref{fig:s1_decoupled} we take the files from the original workflow and we decouple them in the following way: The directory of \texttt{Inputs} with three text files: \texttt{text1.txt, text2.txt,} and \texttt{text3.txt} and the application \texttt{(gnuplot.sh)}. 

\subsubsection{Encapsulating components in containers}
\begin{figure}[h]
\centering
\includegraphics[width=9cm,height=4.5cm]{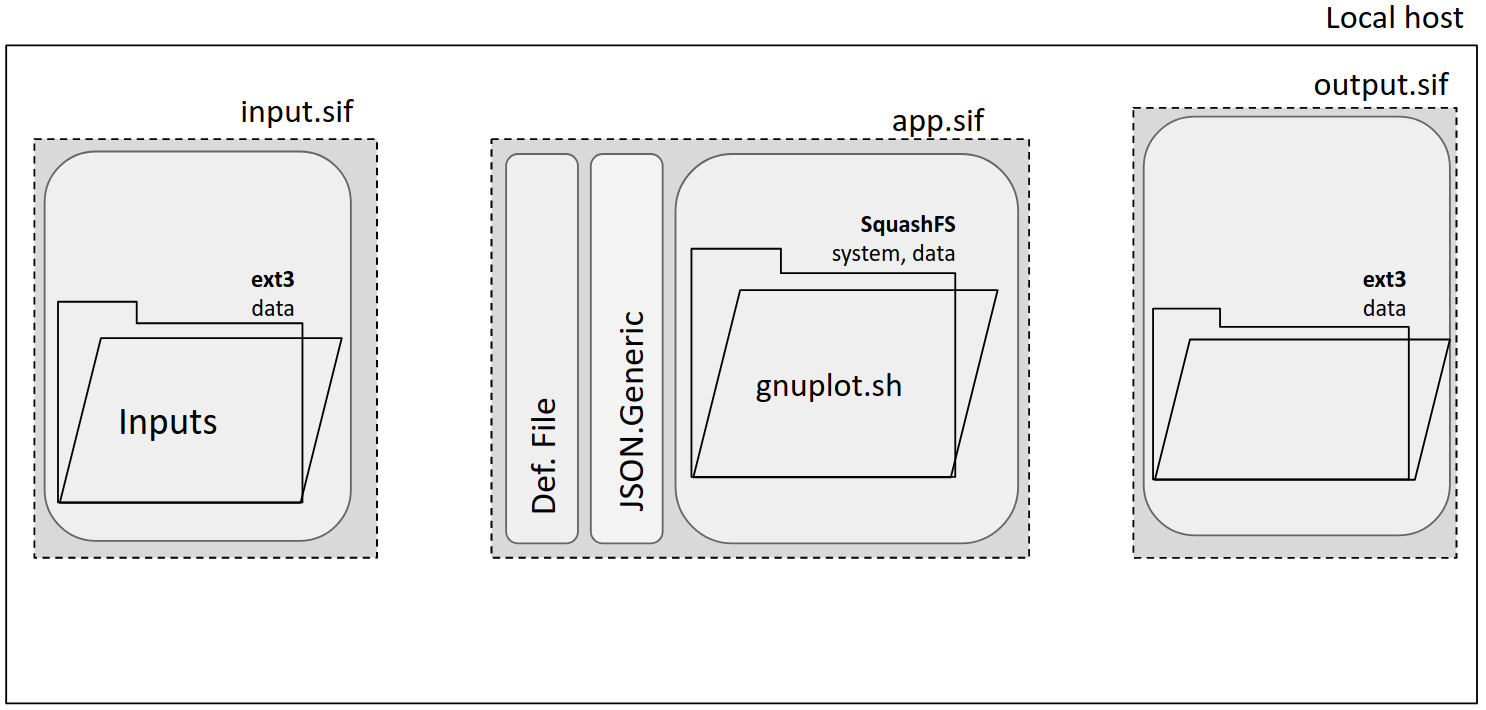}
\caption{Setting up of the containerized workflow components }
 \label{fig:s1_encapsulated}
\end{figure}

In Figure~\ref{fig:s1_encapsulated} we show the set up of our environment with the needed containers for each workflow component. For this specific scenario, our environment has two data containers and one app container.   

All containers have a Singularity Image Format (SIF), which has a layer organizational structure. In each layer you can add a different object. They can be straightforward created and manipulated through the \texttt{singularity sif} tool. 

In terms of the first data container (\texttt{input.sif}),  we use the read and write \texttt{Ext3} file system . It maintains the directory \texttt{Inputs} with the three text files and it is added in a layer of \texttt{input.sif} as a single data partition.

The \texttt{output.sif} is created the same way as \texttt{input.sif}, except that what we compress in the \texttt{Ext3 FS} is an empty directory called \texttt{Outputs}. 

Regarding the application container (\texttt{app.sif}), we build it from a recipe  that defines all the software stack and adds the executable or application. In Listing~\ref{lst:recipe} we present the recipe for scenario 1, which installs Ubuntu 16.04 as the OS and the \texttt{gnuplot} package, and attaches the executable \texttt{gnuplot.sh}. At the end, our \texttt{app.sif} has a single system partition. 

\begin{lstlisting}[caption={Application container recipe.},label={lst:recipe},language=bash]
Bootstrap:docker
From: ubuntu:16.04

%post
apt-get -y upgrade
apt-get -y update && apt-get -y --fix-missing install gnuplot

%files
gnuplot.sh

%runscript
/gnuplot.sh
\end{lstlisting}

\subsubsection{Connecting containers}\label{ssec:implementation_zero-copy}
\begin{figure}[h]
\centering
\includegraphics[width=9cm,height=4.5cm]{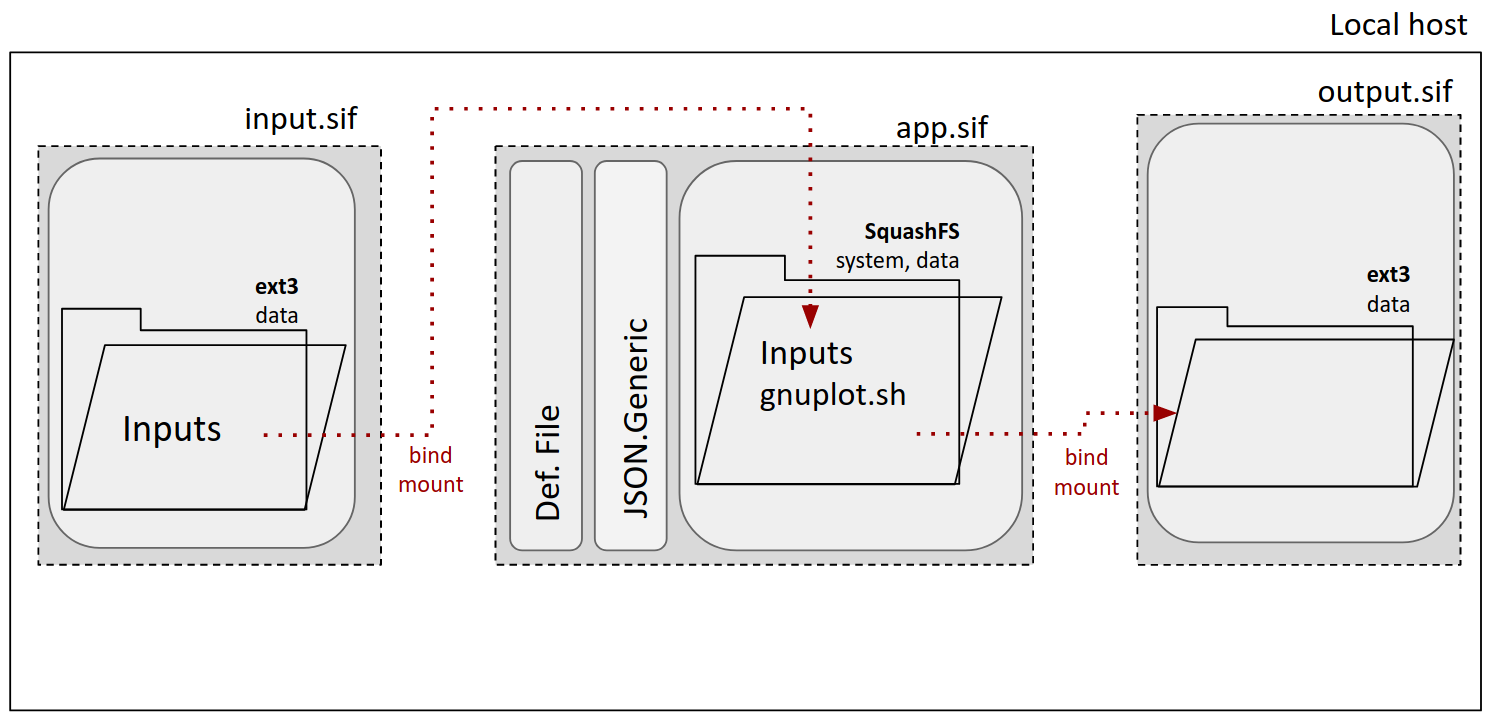}
\caption{Bind mount directories from all the workflow components.}
 \label{fig:s1_connected}
\end{figure}

In Figure~\ref{fig:s1_connected} we connect the containerized components for running the workflow.

 To implement this optimal data transfer we expanded the bind mount feature. A \textbf{bind mount} takes the bind path (a directory) and replicates it under a different point. Previously, the only way to bind mount a path was from anywhere in the host to a container. Now, it is possible to directly bind mount a directory within two different containers. 
 For example, the \texttt{Inputs} directory is bind mounted to a directory inside the application container. Hence, any change in the \texttt{Inputs} directory from the input container is reflected on the app container and vice versa. Same for the app and output containers.

\subsubsection{Executing the containerized workflows}
\begin{figure}[h]
\centering
\includegraphics[width=9cm,height=4.5cm]{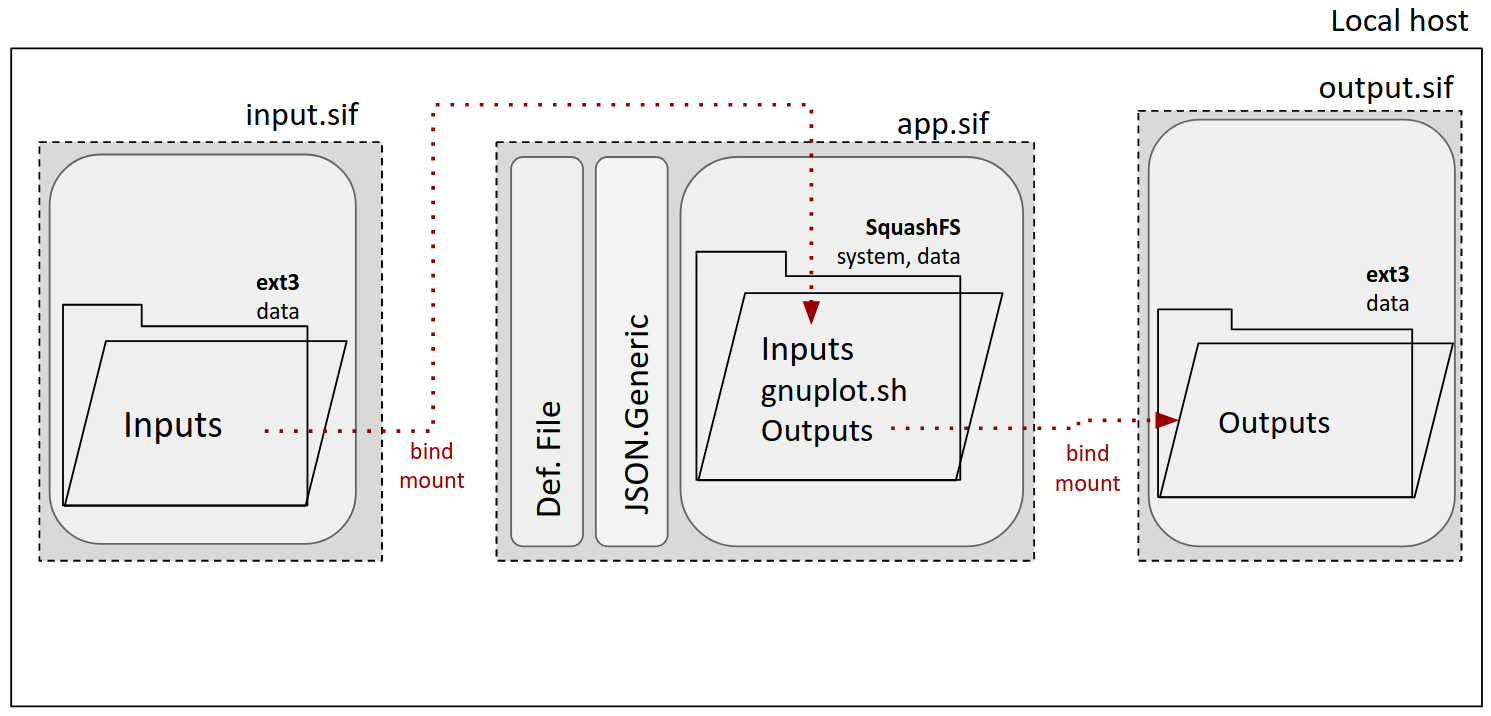}
\caption{Execution of the application and allocation of the results in the output container.}
 \label{fig:s1_executed}
\end{figure}

In Figure~\ref{fig:s1_executed} we present the execution of the containerized workflow. Simple commands from Singularity (\textbf{\texttt{exec}}, \textbf{\texttt{run}} and \textbf{\texttt{shell}}) can be used to run it. After execution we generate the plots: \texttt{plot1.png, plot2.png,} and \texttt{plot3.png}. All the plots are allocated in the \texttt{Outputs} directory within the output container that was already bind mounted to the application container. 

Despite the fact we are hosting the workflow in containers, the users will not need to have a deep knowledge in Singularity to execute our environment. It is as simple as \texttt{singularity exec ./executable container}, or \texttt{singularity run container}, or shell into it (\texttt{singularity shell container}) and from the terminal of the container execute everything as usual.

\subsubsection{Capturing the containers' metadata}
\begin{figure}[h]
\centering
\includegraphics[width=9cm,height=4.5cm]{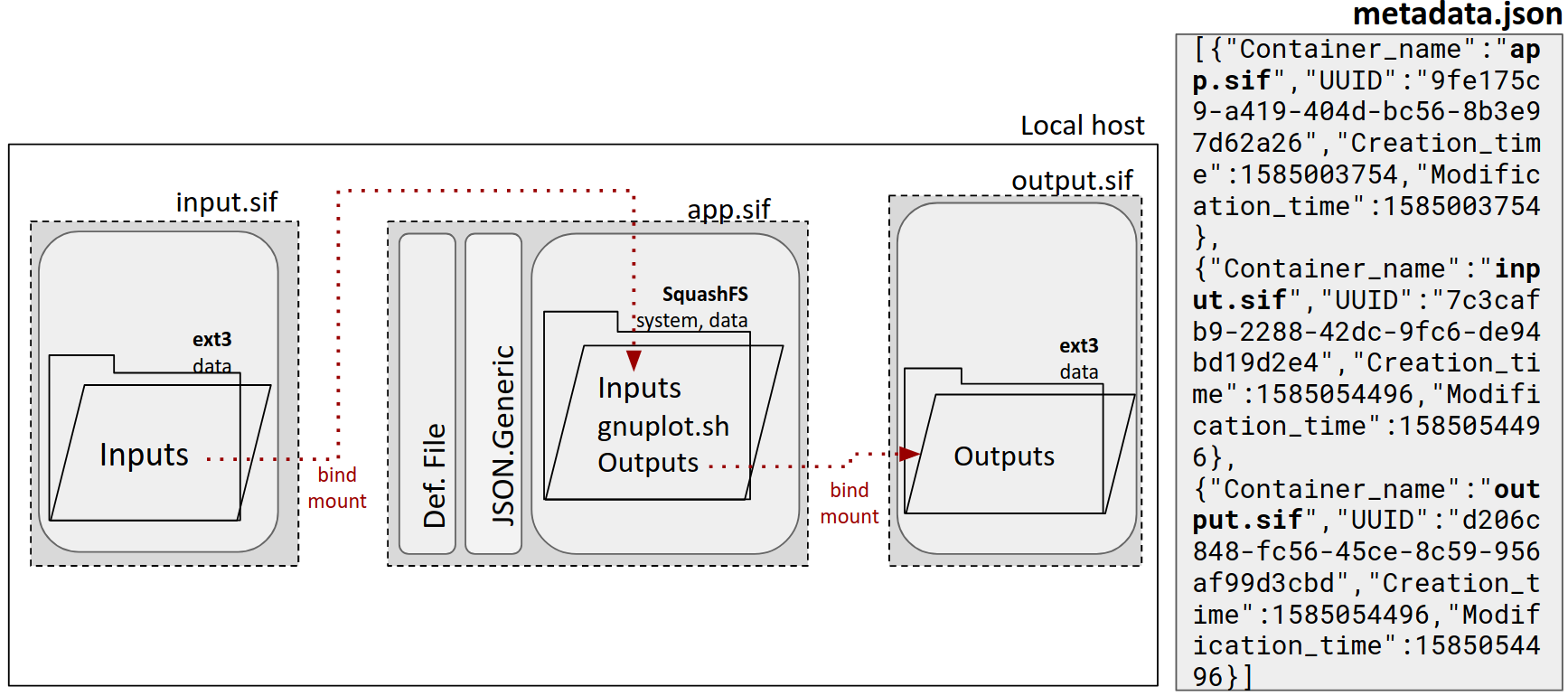}
\caption{Capturing static metadata from all workflow components.}
 \label{fig:s1_captured}
\end{figure}

In Figure~\ref{fig:s1_captured} we show the creation of the metadata that includes important information to identify each of the components from our workflow. Right after execution our Singularity plugin accesses the containers (\texttt{input.sif, appl.sif} and \texttt{output.sif}), and extracts information from their static metadata. This information consists of the name, UUID, creation and modification date. The information is written to the \texttt{metadata.json}  file.  

\subsubsection{Locating the metadata}
\begin{figure}[h]
\centering
\includegraphics[width=9cm,height=4.5cm]{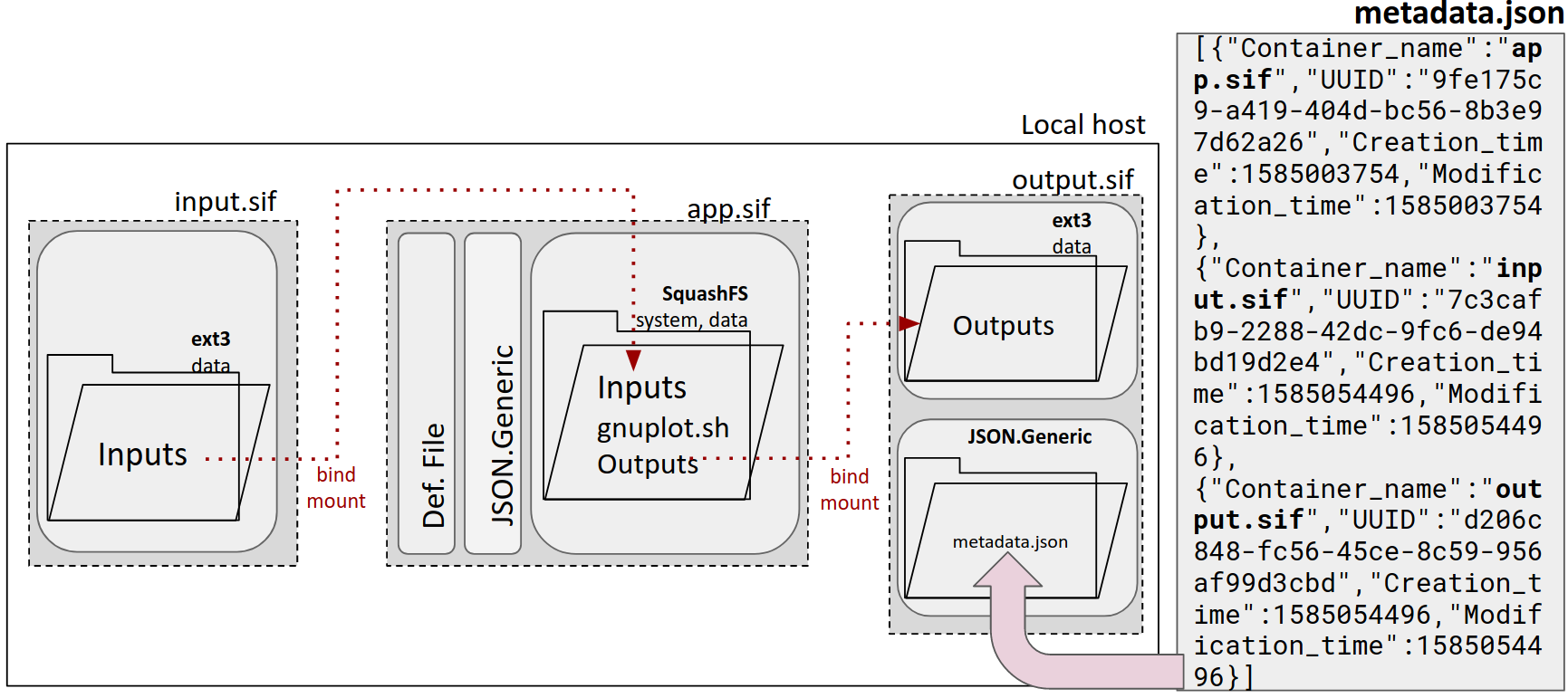}
\caption{Allocation of the metadata as partition of the output data container.}
 \label{fig:s1_collocated}
\end{figure}

Our Singularity plugin also allocates the metadata as a second partition of the output data container as seen in Figure~\ref{fig:s1_collocated}. The type of the partition is a JSON Generic.

\section{Scenario 2: Gnuplot benchmark with multiple inputs}\label{sec:s2}

\begin{figure}[h]
\centering
\includegraphics[width=9cm,height=4.5cm]{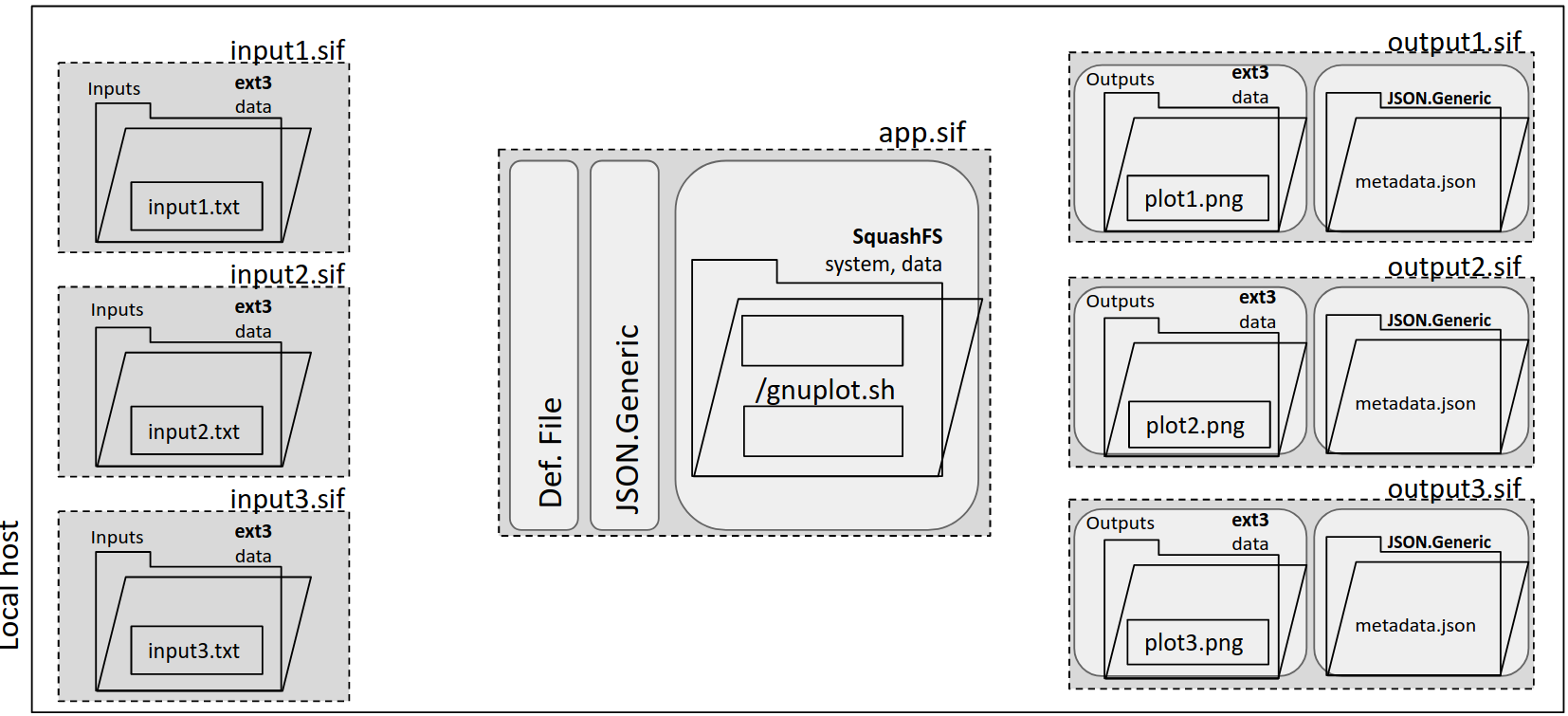}
\caption{The scenario 2 has three inputs (text files) going to the single gnuplot application and generating three plots. }
 \label{fig:s2}
\end{figure}

In Figure~\ref{fig:s2} we present scenario 2, which consists of multiple files as input being modified by a single application generating the respective outputs. We separate each of the three text files from scenario 1 and allocate them in independent containers.
The application is the same from the scenario 1 in Section~\ref{s1}, and the outputs are the final plots on  independent containers. 

The information from each \texttt{metadata.json} differs depending on the components that were used to generate the plot. All metada files have the same \texttt{app.sif} information in their record trail. The difference is the input container that was used to generate that plot. For example, the record trail from \texttt{output1.sif} includes the name, UUID, creation and modification date from \texttt{input1.sif}, \texttt{app.sif} and itself. The same for \texttt{output2.sif} linked to \texttt{input2.sif} and \texttt{app.sif},  and \texttt{output3.sif} with \texttt{input3.sif} and \texttt{app.sif}.  

\section{Scenario 3: Comparison of ML models with shared input}\label{sec:s3}

\begin{figure}[h]
\centering
\includegraphics[width=9cm,height=4.5cm]{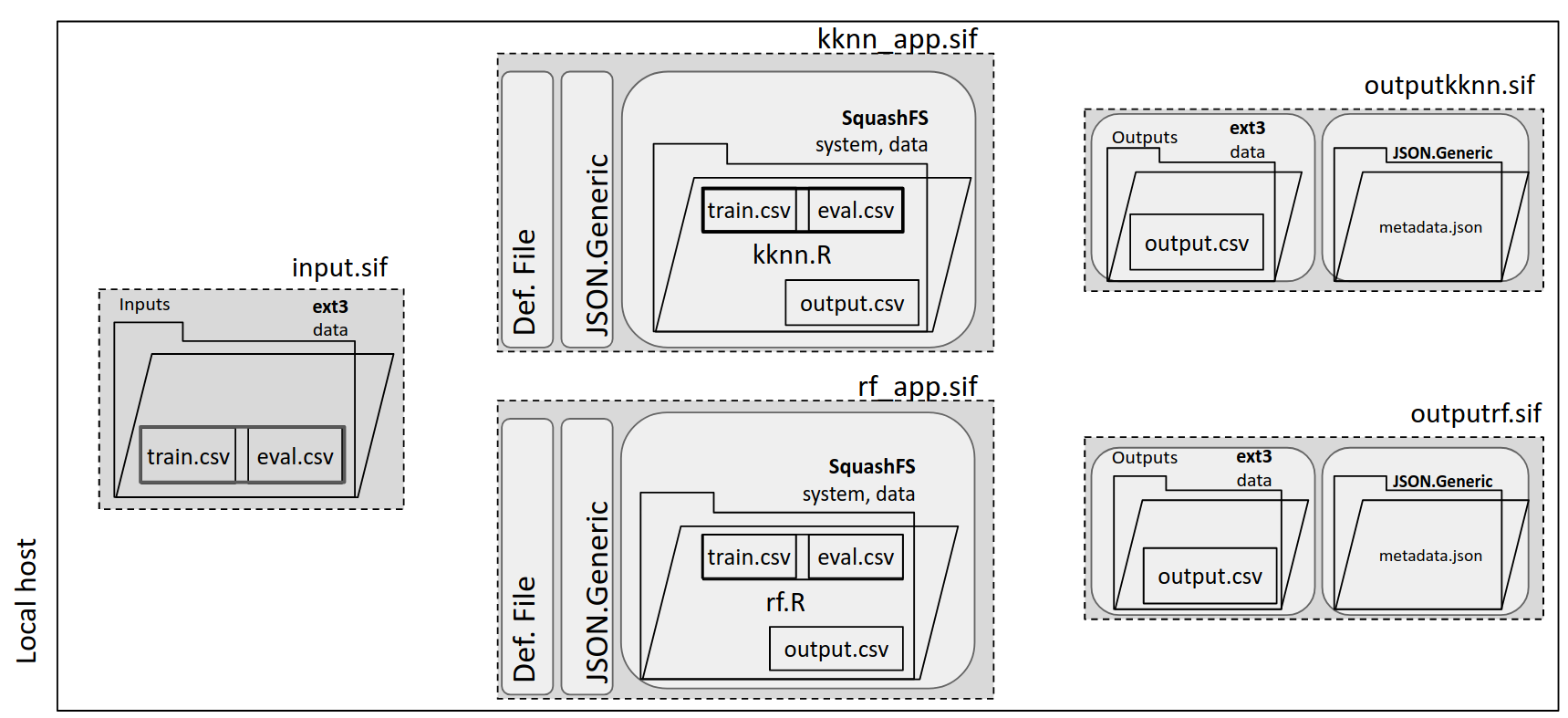}
\caption{The scenario 3 has a input container with two data sets, both going to two app containers and generating two respective outputs containers.}
 \label{fig:s3}
\end{figure}

Scenario 3 presented in Figure~\ref{fig:s3}, consists of a single directory modified by two ML applications (KKNN and random forest) generating the respective predictions. 

The input directory has two data sets: the training and evaluation. The training data contains the coordinates (latitude and longitude) of a region from the United States of America along with some topographic parameters. Each point was taken every $27 km$, that is why there are only $47$ data points (low resolution). The target, that is soil moisture, is also given for each point. The evaluation data set has $44723$ geographic points from the same region because they were taken in a higher resolution ($1km\times1km$) explained by the same features: coordinates and topographic parameters.  
The applications are two ML regression models which objective is to be trained with the first data set (\texttt{train.csv}), and  predict the soil moisture for second data set (\texttt{eval.csv}).  

Both \texttt{metadata.json} files from \texttt{outputkknn.sif} and \texttt{outputrf.sif} include in the record trail the information from \texttt{input.sif}, what changes is the specific application that generated those predictions. Therefore, in the \texttt{outputkknn.sif} we will include the information of \texttt{input.sif} and \texttt{kknn\_app.sif}, and for \texttt{outputrf.sif} we will have it from \texttt{input.sif} and \texttt{rf\_app.sif}.

\section{Scenario 4: Comparison of ML models with split input}\label{sec:s4}

\begin{figure}[h]
\centering
\includegraphics[width=9cm,height=4.5cm]{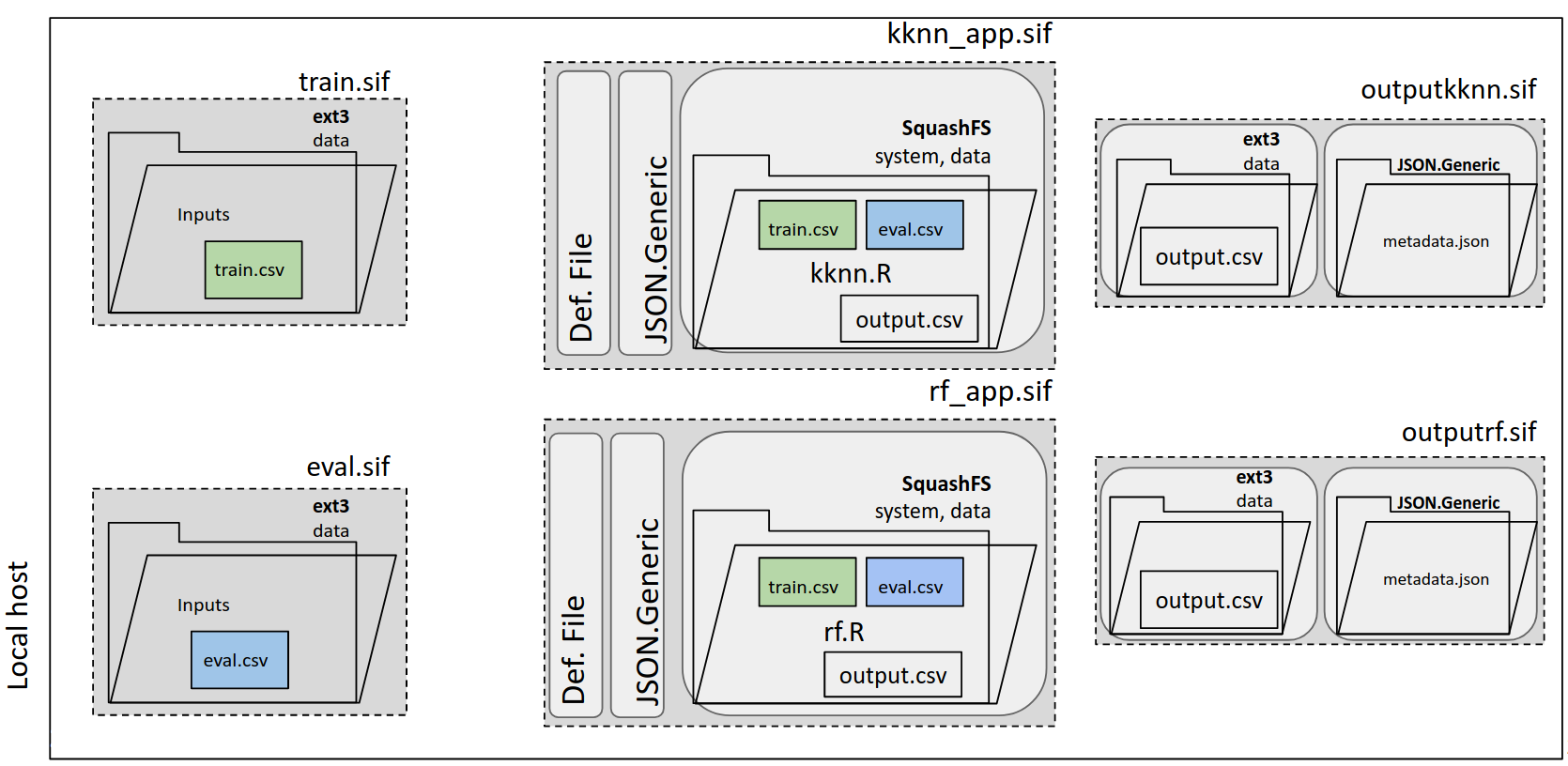}
\caption{The scenario 4 has two input containers, both going to two app containers and generating two respective outputs containers.}
 \label{fig:s4}
\end{figure}

Scenario 4 shown in Figure~\ref{fig:s4} has the same applications as the scenario 3 presented in Section~\ref{sec:s3}, except that each data set (training and evaluation) is allocated in independent containers.

The \texttt{metadata.json} for both output container has the same format as scenario 3, except that now instead of having a single input container (\texttt{input.sif}) we have the information from two: \texttt{train.sif} and \texttt{eval.sif}.

\addtocontents{toc}{\protect\setcounter{tocdepth}{0}}
\section{Summary}
In this Chapter our containerized environment was empirically assessed on four workflows. In Section~\ref{s1} we described the implementation of our application-agnostic containerized environment for the first scenario. The process of implementation is analogous for the other three scenarios. We were able to run two different types of applications without requiring any modification in the code or format. Our environment offers transparent and automatic metadata collection and access. The record trail from the metadata is easy to read, with concise information from the components used during execution. Finally, the metadata keeps a tight connection with the data by being allocated in the same container. 

In the next Chapter~\ref{ch:costs} we will provide the cost in terms of time and space to deploy our containerized environment.
\addtocontents{toc}{\protect\setcounter{tocdepth}{2}}
    \chapter{Effectiveness Containerized Environment} \label{ch:costs}

We evaluate the effectiveness of our environment by measuring its time and space overhead.

\section{Metrics and platform}

For time overhead, we measured the wall-clock time over 500 executions of a given workflow scenario. In this work we consider the more time-consuming ML applications (i.e., scenario 4 in Section~\ref{sec:s4}). 

For space overhead, we mean the memory size of the workflow components (i.e., data and application) when containerized.  In this work we consider the more memory-demanding scenarios 3 in Section~\ref{sec:s3}, and 1 in Section~\ref{s1}. 

We run our tests on a platform with these characteristics: a single socket, 4-core Intel i7-8565U CPU @ 1.8GHz with 16GB RAM and the Linux 5.3.0 operating System. We use the Singularity 3.5 release, which includes the expanded bind mount feature.
\section{Time overhead}

We measure the average wall-clock over 500 executions for the ML applications, specifically, the scenario 4 in Section~\ref{sec:s4}.

This scenario has two inputs and we run two applications, KKNN and RF, as a result we get the predictions from each model. 

\begin{figure}[h]
\centering
\includegraphics[width=1.0\textwidth]{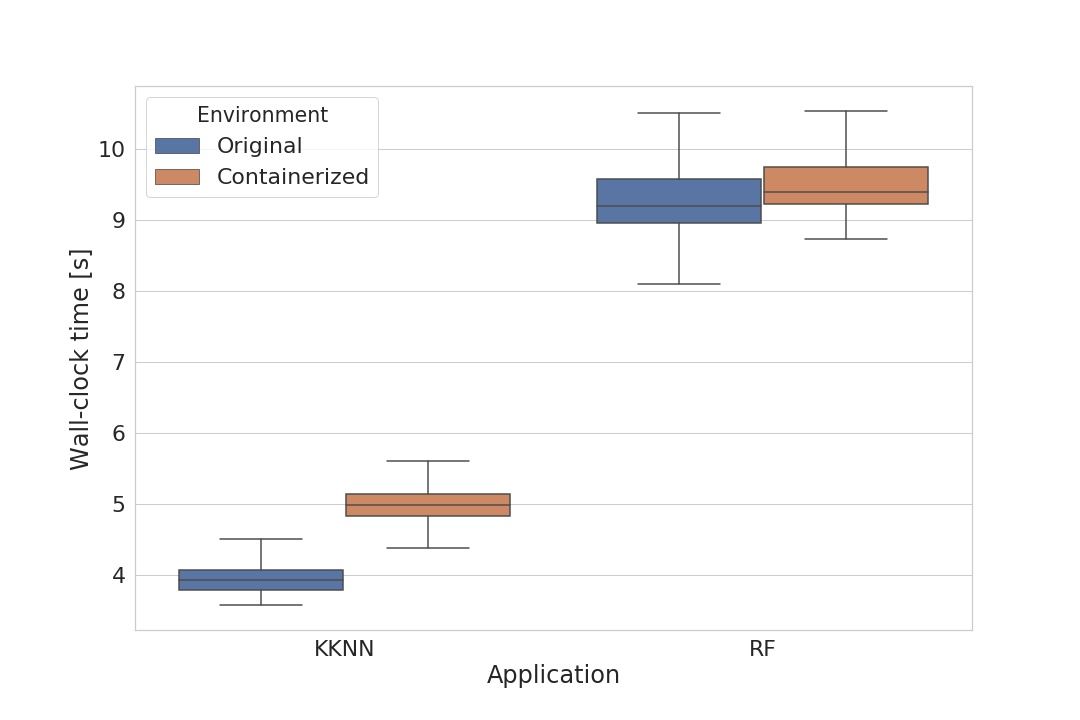}
\caption{Comparison of wall-clock time used during execution between the original and containerized workflow.}
 \label{fig:timeoverhead}
\end{figure}

In the $x$ axis of Figure~\ref{fig:timeoverhead}  we have the type of application, the $y$ axis is the wall clock time. In blue we have the time execution measurements from the original workflow and in orange we have the time taken by our containerized environment.
As we can see for the KKNN the overhead is around $26\%$ and for the RF is $0.7\%$. 

From these results we can conclude that when an application is more complex and it has a larger execution time, like the RF model, the wall-clock time overhead becomes negligible ($0.7\%$). It shows that our environment is promising to keep exploring with real world workflows, such as deep learning workflows or molecular dynamics simulations.  

\section{Space overhead}

We assess both of our containerized components: data and application containers. We compare the size of the files and the executable in the original workflow and after being wrapped in containers. 

\subsection{Data container overhead}
To study the data container space overhead we select the scenario 3 in Section~\ref{sec:s3}. The reason for selecting it is because it has the largest data in one container. As we may recall, the scenario 3  has in its input container the training and evaluation data sets. 

This the composition of our data container: 
\begin{itemize}
    \item Data files  (i.e., Inputs, outputrf.csv, outputkknn.csv)
    \item Workflow metadata (i.e. metadata.json)
\end{itemize}

Table~\ref{tab:datacontainer_overhead} shows the sizes of the data workflow components before and after being containerized. The Input directory has an original size of around $12 MB$ and when it is containerized, it increases to $14 MB$. Concerning the two output containers, they go from $2.2 MB$ to $4.1 MB$. 
Based on these results we can conclude that the space overhead for our tested scenarios is about $2 MB$, which is caused by the file system that we use to wrap the data. 
As discussed in Section~\ref{sec:encapsulate}, for the data container we need a read and write file system. \texttt{EXT3} FS is the one available in Singularity that fits these requirements.

\texttt{EXT3} FS needs $2 MB$ of extra space for specific file system information. In this sense the data container space overhead is directly related with the file system used, consequently if Singularity adds other file systems, we can explore them and reduce the overhead. 

\begin{table}[!ht]
\centering
\caption{Data container space overhead measured for scenario 4.}
\begin{tabular}{|c|cccc}

 \hline
 \multirow{2}{*}{\textbf{Scenario}} & \multicolumn{2}{c|}{\textbf{Original Workflow}} & \multicolumn{2}{c|}{ \textbf{Containerized Workflow}} \\ \cline{2-5}
 & \multicolumn{1}{c|}{File} & \multicolumn{1}{c|}{Size [KB]} & 
 \multicolumn{1}{c|}{File} & \multicolumn{1}{c|}{Size [KB]} \\
 
 \hline
 \multirow{4}{*}{S3} 
 & \multicolumn{1}{l|}{Inputs} & \multicolumn{1}{l|}{$12071$} & \multicolumn{1}{l|}{input.sif} & \multicolumn{1}{l|}{$14368$} \\
 
 & \multicolumn{1}{l|}{outputkknn.csv} & \multicolumn{1}{l|}{$2264$} & \multicolumn{1}{l|}{kknn\_output.sif} & \multicolumn{1}{l|}{$4132$} \\ 
 
 & \multicolumn{1}{l|}{outputrf.csv} & \multicolumn{1}{l|}{$2264$} & \multicolumn{1}{l|}{rf\_output.sif} & \multicolumn{1}{l|}{$4132$} \\
 \hline
\end{tabular}
\label{tab:datacontainer_overhead}
\end{table}

\subsection{Application container overhead}
To study the application container space overhead we select the scenario 1 in Section~\ref{s1}. This scenario consists of a single input, application and output. We choose this scenario because it has the basic application container composition. It only requires the OS and one package, \texttt{gnuplot}. Scenario in Section~\ref{sec:s2} has the same application container. The simplicity of the software setting from the application container allows the analysis of the overhead produced by those dependencies that make it possible to be a container. 

This is the composition of our application container: 
\begin{itemize}
    \item Application executable  (i.e., gnuplot.sh)
    \item SW package includes system tools, system libraries and settings (i.e., Ubuntu 16.04, gnuplot)
\end{itemize}

Table~\ref{tab:appcontainer_overhead} shows the size of the application workflow component before and after being containerized.
\begin{table}[!ht]
\centering
\caption{Application container space overhead measured for scenario 1.}
\begin{tabular}{|c|cccc}

 \hline
 \multirow{2}{*}{\textbf{Scenario}} & \multicolumn{2}{c|}{\textbf{Original Workflow}} & \multicolumn{2}{c|}{ \textbf{Containerized Workflow}} \\ \cline{2-5}
 & \multicolumn{1}{c|}{File} & \multicolumn{1}{c|}{Size [KB]} & 
 \multicolumn{1}{c|}{File} & \multicolumn{1}{c|}{Size [KB]} \\
 
 \hline
 \multirow{4}{*}{S1} 
 & \multicolumn{1}{l|}{gnuplot.sh} & \multicolumn{1}{l|}{$4$} & \multicolumn{1}{l|}{gnuplot.sh} & \multicolumn{1}{l|}{$4$} \\
 
 & \multicolumn{1}{l|}{gnuplot} & \multicolumn{1}{l|}{$139$} & \multicolumn{1}{l|}{gnuplot} & \multicolumn{1}{l|}{$139$} \\ 
 
 & \multicolumn{1}{l|}{Software dependencies} & \multicolumn{1}{l|}{$>40000$} & \multicolumn{1}{l|}{Software dependencies} & \multicolumn{1}{l|}{$153000$} \\
 \hline
\end{tabular}
\label{tab:appcontainer_overhead}
\end{table}

As defined earlier, the application container packages the software dependencies. Hence, to make a fair comparison we can not compare only the executable with the app container. That is why, we compare the executable, package gnuplot, and the O.S. and other software dependencies. 

The executable (\texttt{gnuplot.sh}) has the same size in the original and containerized workflow. The package gnuplot plot as well, $139 KB$ in both cases.
What changes are the rest of the software dependencies, for example, in scenario 1, Ubuntu 16.04 was selected as  the Linux distribution. This distribution can be as small as $40 MB$, depending on all software that the user requires to add. After containerizing all the software dependencies, we also include the software settings that make possible to have a container, which leads to a size of $153 MB$.

From these results, we can conclude that our environment identifies and packages the software stack and its dependencies. This property ensures the replicability, reproducibility, and transparency of the software system. The reason our environment ensures them, is that we guarantee the user to always have the same versions of OS, libraries and packages, no matter on which platform we execute the workflow. 

\addtocontents{toc}{\protect\setcounter{tocdepth}{0}}
\section{Summary}
In this Chapter we provided the cost in terms of time and space to deploy our containerized environment. Regarding the time, we measured the wall clock time over 500 executions of scenario 4 presented in Section~\ref{sec:s4}, before and after containerization. Based on this test, we concluded that with more complex and larger applications like the RF model, the wall-clock time overhead becomes negligible ($0.7\%$). For the space overhead, we compared the size of the workflow components (data and application) before and after being containerized for scenario 1 (Section~\ref{s1}) and scenario 3 (Section~\ref{sec:s3}). Based on this test, we concluded that our environment identifies and packages the software stack and its dependencies, guaranteeing to have a replicable, reproducible and transparent software system no matter on which platform the workflow is executed. Another conclusion is that the space overhead of the data and application containers is linked to the selection of OS, software stack and filesystem.

In Chapter~\ref{ch:conclusion} we will summarize the benefits of executing a workflow under our application-agnostic containerized environment. Besides, we will recapitulate the considerations and the lessons learned by designing, building and implementing it. Finally, we will state the future work to expand our environment.     
\addtocontents{toc}{\protect\setcounter{tocdepth}{2}}

    \chapter{Conclusions and Future Work} \label{ch:conclusion}

\section{Summary} \label{sec:summary}
We adopted the PASS\cite{pass} approach by capturing metadata at OS-level, and  then we enhanced it by leveraging cutting edge container technologies. We built a first prototype of an application-agnostic containerized environment which does not require modification of the format or code of the application. It features zero-copy data transfer between containers. It builds the record trail of different scenarios for simple workflow metadata. And it has consistency and a tight connection between the data and its metadata.

We identified two container limitations. The first limitation was the lack of a direct data transfer between containers. We addressed this limitation by expanding the bind mount feature, so now direct paths inside containers can be connected. The second limitation was the unavailability of support for building the workflow record trail. We addressed the second limitation by developing a container plugin that captures metadata.

We tested the environment on four workflows with production-style applications, such as, visualization and machine learning models. 
Finally, we evaluated the cost in terms of time and space of running an application in our containerized environment. The overhead execution time introduced by our environment is $26\%$ for the KKNN model in scenario 4. However, when the application is more complex and with a larger execution time, like the RF model, the overhead reduces to a $0.7\%$. We identified that our environment is able to reproduce and track our workflow from a OS-level, by packaging all the software dependencies in the application container. We found that the space overhead for data and application container is driven by the OS, software stack and file system selection.

\subsection{Compendium of supported functionalities}
In Table~\ref{tab:functionalities} we summarize and highlight the functionalities of our environment that ensure replicability, reproducibility, transparency and traceability.
\begin{table}[]
\centering
\caption{Functionalities from our environment that ensure the replicability, reproducibility and traceability of scientific workflows.}
\resizebox{\textwidth}{!}{
\renewcommand{\arraystretch}{1.7}
\begin{tabularx}{\textwidth}{p{.25cm}>{\raggedright\arraybackslash}p{4cm}X}
\toprule[2pt]
\multicolumn{2}{l}{\textbf{Supported functionality}} &  \textbf{Description} \\%
    \midrule[2pt]
    \multicolumn{3}{l}{\textbf{Replicability and reproducibility}} \\%
    & Automatic workflow metadata  & Automatic documentation of inputs, applications,  outputs and command line after each execution \\%
     & Unique identification of components & Encapsulation of the files in containers; Universally unique identifiers for each container  \\%
    & Packaging software system & Application container includes the software package which has all the system tools, libraries and system-settings \\%
    \midrule[1pt]
    \multicolumn{2}{l}{\textbf{Transparency}}            &  \\%
    & Simple tool integration & Containerized framework; no modifications are required on the application to run it on our framework \\%
    & Programmatic metadata access  &  Singularity \texttt{sif} tool enables straightforward manipulation of each container \\%
    & Easy-to-read record trail & Concise information of the components used in execution in JSON format (name, UUID, creation and modification date) \\%
    \midrule[1pt] 
    \multicolumn{2}{l}{\textbf{Traceability}}            &  \\%
    & Integration of data and metadata & Tight relation between the data and its metadata by collocating the metadata file as a partition of the output data container \\%
    & Record trail assembly  & We capture the interaction between components inside the workflow (data and apps), and their identifications \\%
    & Unique identification of components & Encapsulation of the files in containers; Universally unique identifiers for each container \\%
\bottomrule[2pt]
\end{tabularx}
}
    \label{tab:functionalities}
\end{table}

We ensure \textbf{replicability} and \textbf{reproducibility} by automatic collection of workflow metadata, a unique identification per each workflow component, and packaging the software system in the application container. We ensure \textbf{transparency} by having a containerized framework, a programmatic metadata access, and an easy-to-read record trail. And finally, we are able to \textbf{trace} the execution history of new data and every workflow component, by integrating the data and the metadata, assembling the record trail of data moving within the workflow, and uniquely identifying each component.

\section{Future work} \label{sec:future}

In the future we intend to generalize the metadata specifications to target a broader range of workflows, to include more software characteristics relevant for those workflows, and add more detailed information from workflow components. 
We aim to catalogue record trail for a broad range of workflows from: single to multiple applications, moderate (single-node) to large resource requirements (multi-nodes) and compute-intensive or data-intensive, or both.
We desire to include system software characteristics relevant for those workflows, such as, performance metrics or even platform-related information. 
Finally, we want to add fine-grained information from the workflow components, an example is to separate input deck from the application. 

Based on the experience, we know we will need to expand the range of available container features. For example, in order to automate the setup of our containerized environment, we could allow the creation of containers at run time. We also need to assure security of containerized environments by exploring container properties, like encryption and immutability. Moreover, we want to be able to reproduce and replicate a workflow by retrieving the record trail and the workflow containers. To do this, we could implement another plugin which would extract the UUIDs from the metadata and with the command line, re-assembly and re-execute the workflow. 

Finally, we will need to augment the system software support. To allow run time container creation we would have to modify the kernel. To support large data containers and programmatic access to the metadata, we would need to modify the file system. And to support metadata for in memory objects and creation of containers at run time due system memory call, we would require memory system modifications.

    \makeBibliographyPage 
\newpage


\utbiblio{#1}{IEEEtran}{references-dissertation}

    
    \makeAppendixPage{2}   
    \appendix    
    
\section{Evaluation of data transfer between containers}

In this appendix we will evaluate the zero-copy data transfer method.

Since Singularity lacked a method to transfer directly data between containers, together with the Singularity team, the the bind mount we worked with the Singularity team to expand the bind mount feature. Now instead of only mapping directories from host to container, we are able to link directories within two different containers, as explained in Section~\ref{ssec:implementation_zero-copy}.

The two-copy data transfer, moves data from one container to another always going through host.On the other hand zero-copy data transfer replicates the directory from one container to a path in the second container.

To evaluate the benefits of this feature we compare the time taken to transfer data between containers for each method. We use the configuration from scenario 1 in Section~\ref{s1}, and the VQA\cite{VQAdataset} dataset. 

We compare the wall-clock time that takes to copy data from the input container to the output container passing through the application container.
We select the data set from VQA which is a new data set containing open-ended questions about images. It includes 82,783 images of 81 types of objects and it weights $14 GB$.

In Listing~\ref{lst:2-copy} we can see how we moved the data from the input container to output container. First, we move from input container \texttt{input.sif} to a directory un the host system by using \texttt{sif dump}. Then, we bind mount two directories, the \texttt{Inputs} and \texttt{Outputs} directories in the host to the application container (\texttt{app.sif}). Within the application container, we move the data from \texttt{Inputs} to \texttt{Outputs}. Finally, the data in \texttt{Outputs} in the host is added as a data partition in the output container \texttt{output.sif}. In the end of our transfer, our data file is in the input container and output container.
\begin{lstlisting}[caption={two-copy data transfer between containers.},label={lst:2-copy},basicstyle=\scriptsize, language=bash]
singularity sif dump 1 input.sif > Inputs/Inputs.ext3

singularity exec --contain -B Inputs -B Outputs app.sif mv Inputs/* Outputs/.

singularity  sif add --datatype 4 --partarch 2 --partfs 2 --parttype 3 output.sif Outputs/Inputs.ext3
\end{lstlisting}

In zero copy, as seen in Listing~\ref{lst:0-copy}, we directly connect the directories from the input and output containers to the app container. We execute within \texttt{app.sif} to copy the data between directories.
\begin{lstlisting}[caption={zero-copy data transfer between containers.},label={lst:0-copy},basicstyle=\scriptsize,language=bash]
singularity exec --contain -B input.sif:/Inputs:image-src=/Inputs -B output.sif:/Outputs:image-src=/Outputs app.sif cp -r /Inputs/* Outputs/.
\end{lstlisting}

We run each process 50 times and measure wall clock time. We run our tests on a platform with these characteristics: a single socket, 4-core Intel i7-8565U CPU @ 1.8GHz with 16GB RAM and the Linux 5.3.0 operating System. We use the Singularity 3.5 release, which includes the expanded bind mount feature.

\begin{figure}[h]
\centering
\includegraphics[height=8cm]{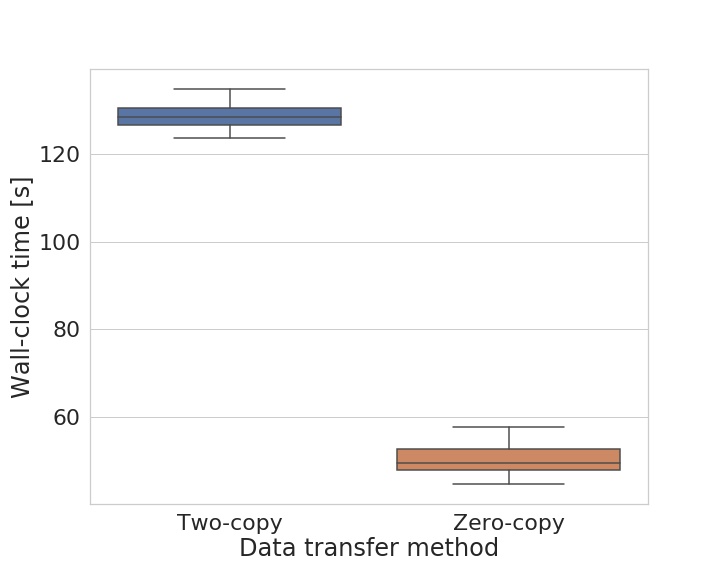}
\caption{Comparison of wall-clock time used to copy data from input to output container between the two-copy and zero-copy data transfer method.}
 \label{fig:datatransfer}
\end{figure}

 In Figure~\ref{fig:datatransfer} we present the wall-clock time speed-up from two-copy to zero-copy data transfer between containers. On the $x$ axis is the data transfer method, on the $y$ we have the wall clock time and we present the distribution for each method. We obtain a average speed-up of $60.94\%$. This assures that the zero-copy data transfer method has significantly better performance than the previous implementation from Singularity. 
    \addToTOC{Vita}
    \chapter*{Vita} \label{ch:vita}
Paula Olaya is an M.S - Ph.D. student in Computer Science advised by Dr. Michela Taufer at the University of Tennessee, Knoxville. Her expected M.S. graduation date is in Spring 2020. She has a B.S. in Electronics Engineering from Pontifical Xavierian University in Colombia, with an emphasis on bio-signals and signal processing. Olaya's research interests in high-performance computing include the creation of general application-agnostic containerized environments to allow intrinsic replicability, reproducibility, transparency, and traceability of scientific workflows. She has been involved in projects with HPC and machine learning applications, also benchmarking and analyzing performance in terms of time and power usage for FFTs and sparse matrix solvers at architecture and algorithm level for an astrophysics application.
\end{document}